\documentclass{article}
\usepackage[english]{babel}
\usepackage{graphicx}
\usepackage[all]{xy}
\usepackage{textcomp}
\usepackage{tipa}
\usepackage{color}
\usepackage{amsmath}
\usepackage{stackrel}
\usepackage{amssymb}
\newcommand{\be}{\begin{equation}}
\newcommand{\ee}{\end{equation}}
\newcommand{\ben}{\begin{eqnarray}}
\newcommand{\een}{\end{eqnarray}}
\newcommand{\bes}{\begin{subequations}}
\newcommand{\ees}{\end{subequations}}

\begin{document}
\title{{\bf ${\bf {\cal N}=2}$ Supersymmetric quantum mechanics of \\ ${\bf N}$ Lieb-Liniger-Yang bosons on a line}}

\author{J. Mateos Guilarte$^{(a)}$ and  A. Moreno Mosquera$^{(b)}$
\\ {\normalsize {\it $^{(a)}$ Departamento de Fisica
Fundamental}, {\it University of Salamanca, SPAIN}}\\ {\normalsize {\it $^{(b)}$Facultad Tecnol\'{o}gica. Universidad Distrital Francisco Jos\'e de Caldas, Bogot\'a, COLOMBIA.}}}

\maketitle


\begin{abstract}
A supersymmetric generalization of the Lieb-Liniger-Yang dynamics governing $N$ massive bosons moving on a line with delta interactions among them at coinciding points is developed. The analysis of the delicate balance between integrability and supersymmetry, starting from the exactly solvable non supersymmetric LLY system, is one of the paper main concerns. Two extreme regimes of the $N$ parameter are explored: 1) For few bosons we fall in the realm of supersymmetric quantum mechanics with a short number of degrees of freedom, e.g., the SUSY P$\ddot{\rm o}$sch-Teller potentials if $N=1$. 2) For large $N$ we deal with supersymmetric extensions of many body systems in the thermodynamic limit akin, e.g., to the supersymmetric Calogero-Sutherland systems. Emphasis will be put in the investigation of the ground state structure of these  quantum mechanical systems
enjoying ${\cal N}=2$ extended supersymmetry without spoiling integrability. The decision about wether or not supersymmetry is spontaneously broken, a central question in SUSY quantum mechanics determined from the ground state structure, is another goal of the paper.
\end{abstract}


\section{Introduction}

Literature on $N$ particles moving back and forth a line solvable models is quite extensive and covers systems in which different types of interaction and boundary conditions are considered. Bethe's work \cite{Bethe} on Heisenberg's isotropic spin chains constitutes the first meaningful example of integrable many body quantum system. Bethe presented his famous ansatz in order to obtain the eigenstates of energy corresponding to the Hamiltonian of the Heisenberg model that describes a fixed-spins chain in one-dimensional network places. In that model, Bethe implemented his ansatz keeping in mind that for $N$ particles systems, the configuration space can be divided into $N!$ regions, each corresponding to the different arrangements of the particles over the line from left to right. Bethe's approximation implies that the wave function in each region is a superposition of plane waves as if the particles did not interact in such a way that when applying matching conditions to the wave function in the interfaces of these regions he obtained a system of equations for the coefficients and wave numbers for plane waves, called Bethe equations. Later, Hulthen \cite{Hulthen}  used the same ansatz to describe the anti-ferromagnetic case of the Heinsenberg's model. We would have to wait another two decades for a generalization of Bethe's ansatz to be implemented by Orbach \cite{Orbach}  and Walker \cite{Walker}, then later more accurately by Yang and Yang \cite{YYan,YYang}, in their study of XXZ spins chain. In 1963, Lieb and Liniger \cite{Lieb,Liniger} successfully applied Bethe's ansatz to solve a bosonic $N$ particle system moving freely along the real line except at coinciding points, where two particles interact via a Dirac delta-type potential. The Lieb-Liniger system was supplemented by periodic boundary conditions if the chain of particles is finite. Later on, McGuire \cite{McGuire} examined scattering boundary conditions when the particles are allowed to move over the whole real line, discovering the scattering wave functions. These and other works proved a fundamental property of the integrable $N$ particle systems: the scattering of $N$ particles can be factorized as the product of two-particle scattering in such a way that an invariance exists regarding the order in which these processes take place. In Reference \cite{Yang}, Yang developed the generalization of these models to the case of fermionic particles using a variant of the Bethe ansatz and wrote the celebrated Yang-Baxter equations as the condition of internal consistency for the factorization of the scattering of $N$ particles in two -particle processes. Apart from the marking, Baxter obtained the same equations in his study of the XXZ model, which led him to establish the connection between one-dimensional many body systems in Quantum Mechanics and perfectly solvable two-dimensional classical systems in Statistical Mechanics \cite{Baxt,Baxte,Baxter,Sutherland}. The Yang-Baxter equations have an important r$\hat{\rm o}$le connecting many areas of Mathematical Physics, Statistical Mechanics and Mathematics. In that respect, it has been demonstrated that the knot invariants can be built from lattice models in Statistical Mechanics \cite{Wu,Wadati}, whereas in Mathematical Physics, the same equations combined with the concept of transfer matrix are fundamental for the study of the \lq\lq Quantum Inverse Scattering Method\rq\rq, a procedure developed by Faddeev, Sklyanin, Takhtajan and their collaborators \cite{Sklyanin,ESklyanin} that generalizes to the quantum world the classical Inverse Scattering method of solving many integrable ODE's and PDE's equations. The key ingredient, the Bethe ansatz, has been successfully applied to models in Statistical Mechanics and Quantum Field Theory in $(1+1)$-dimensional space-time \cite{Zamolodchikov}.

The issue of the extension of some of these $N$-body integrable systems to the framework of supersymmetric quantum mechanics was addressed
at the beginning of the nineties of the past century, see e.g. References \cite{Freedman}-\cite{Brink0}-\cite{Shastry}, specifically over Calogero-Sutherland models of several types \cite{Hallnas}-\cite{Sutherland}-\cite{Calogero0}-\cite{Calogero}. The interplay between
integrability and supersymmetry in this research is very intriguing and it was clarified to a good extent in the References \cite{Brink}
and \cite{Desrosier} relying on the Lie algebras hidden in the Calogero-Moser-Sutherland models and their representations. Another works in a similar vein where
Calogero models of Type V Reference \cite{Manton}, pair potentials of the class of isotonic oscillators, are extended to supersymmetric quantum mechanics is developed in Reference \cite{Manton}.

The main theme in this paper is the construction of a ${\cal N}=2$ supersymmetric quantum mechanical model{\footnote{A very good compendium of the structure of Supersymmetric Quantum Mechanics may be found e.g. in Reference \cite{Freund}.}} extending the Lieb-Liniger-Yang ${\cal N}=0$ system describing the dynamics of $N$ bosons in a line interacting pair-wise via Dirac $\delta$ potentials at coinciding points, see \cite{Lieb}-\cite{Liniger}-\cite{Yang}.
 We shall follow the structure developed by Wipf et al. in Reference \cite{Wipf}. A brief summary of ${\cal N}=2$ supersymmetric quantum mechanics extending Hamiltonian quantum systems with $N$ degrees of freedom to a  SUSY framework is as follows:
\begin{enumerate} \item Each \lq\lq bosonic\rq\rq operator, typically coordinates labeling the particle position in ${\mathbb R}^N$ is suplemented with a \lq\lq fermionic\rq\rq operator. The fermionic operators are represented by elements of
the Euclidean Clifford algebra of $\mathbb{R}^{2N}$.

\item The space of states in ${\cal N}=2$ supersymmetric quantum mechanics is the tensor product of the quantum particle Hilbert space $L^2({\mathbb R}^N)$ times the Fermionic Fock space of dimension $2^N$. A ket basis in this enlarged space of states can be wriiten as
\[
\vert x_1,\alpha_1; x_2,\alpha_2; \cdots \cdots . x_N, \alpha_N \rangle \, \, \, , \, \, \,  (-1)^{\alpha_j}=\pm 1 \, \, , \, \, \forall \, j=1,2, \cdots N \, ,
\]
meaning that, besides the particle position, there are a Fermionic state for each coordinate that may be empty, if $\alpha_j=0$, or occupied, when $\alpha_j=1$. Thus, the states of the basis are clasified as irreducible representaions of $SO(N)$ of dimension $\binom{N}{j}$.

\item The third, and perhaps main, ingredient are the generators of the supersymmetry algebra or supercharges. The supercharge are first-order in the momenta dressed with the Fermionic operators plus the gradient of a function called the superpotential. The whole construction
produces a supersymmetric Hamiltonian which is factorized in terms of the supercharge operator and its adjoint being thus the super-Hamiltonian invariant within
the supersymmetry algebra and the supercharges themselves constants of motion.
\end{enumerate}

 Searching for building a supersymmetric extension of a given quantum mechanical problem the crucial step is the choice of superpotentiall because in many cases there are different possibilities {\footnote { The ambiguity in choosing different superpotentials leading to the same non supersymmetric potential is discussed in Reference \cite{Matgui}.}}.
Wipf and his collaborators performed a very interesting promotion of the Hydrogen atom in $N$ dimensions to its generalization in ${\cal N}=2$ extended supersymmetric Quantum Mechanics. These authors introduced the supersymmetric Coulomb potential by means of an shrewd choice of \lq\lq superpotential\rq\rq. Their choice of superpotential meant that, in the scalar sectors where all the Fermionic states are either empty or occupied, the interactions emerging were either attractive or repulsive Coulomb potentials plus a constant ensuring that the ground state energy is zero. Moreover, the non dynamical $SO(N+1)$ symmetry of the non-SUSY Coulomb problem was preserved in the supersymmetric generalization proposed in \cite{Wipf} and the promotion of the Runge-Lenz vector to a supersymmetric invariant operator allowed these authors to solve algebraically the spectral problem also in the supersymmetric Hydrogen atom. The main merit of the Wipf et al. supersymmetric formulation of the Coulomb problem is thus preservation of super-integrability.

The generators of the $SO(N)$ group transformations acting on the $N$-particle ${\mathbb R}^N$ configuration space are extended to act on the whole Fermionic space by means of the fermionic operators. The fermionic degrees of freedom respond thus to the existence
of an \lq\lq internal\rq\rq spin structure albeit restricted to spin integer states. The extra interactions induced by the Wipf et al.
supersymmetric construction over the standard Hydrogen atom obeys to quadrupolar forces arising between states carrying non zero intrnal spin in the matricial sectors. In the scalar sectors, however. which belong to the trivial representation of $SO(N)$ the interactions are
purely monopolar Coulomb, either repulsive or attractive, potentials.

We confront a similar endeavour: starting from the LLY system of $N$ bosonic particles we want to build around this model a supersymmetric structure. The choice of superpotential in such a way that pair-wise $\delta$-potentials arise in the supersymmetric LLY Hamiltonian presents several alternatives. Our guiding principle in this choice is the persistence of the
applicability of the Bethe ansatz. In fact, given that the supersymmetric Hamiltonian in the scalar sectors reduces to the non SUSY LLY Hamiltonian we shall be able of identifying eigenfunctions in nearest sectors to the scalar ones by means of the action of the supercharge operators $\hat{Q}^\dagger$ and $\hat{Q}$. We remark that the same supersymmetric generalization of the $SO(N)$ group of transformations exists in our Lieb-Liniger-Yang supersymmetric model, allowing us to interpret the wave functions in the matricial sectors as due to the existence of an internal spin degree of freedom. Moreover, the extra interactions included in the supersymmetric structure maens that there are contact interactions also bewtween states of different spin within an irreducible representation. It must be stressed thatneither
the bosonic nor the fermionic $SO(N)$ generators give rise to symmetries of the non-SUSY or SUSY LLY models. 
We are interested in knowing the ground state structure because it is crucial to determine whether supersymmetry is spontaneously broken. Knowledge of one-particle supersymmetric Hamiltonians encompassing Dirac $\delta$-potentials, see Reference \cite{Mosquera1}, will help us in the analysis of the spectra of Lieb-Liniger-Yang supersymmetric Hamiltonians.

The organization of the paper is as follows: In Section \S.2 a summary of the spectrum of the LLY Hamiltonian is offered. We describe first the scattering eigenfunctions responding to two-particle exchanges when the $\delta$-interactions are repulsive. We also consider the case of attractive $\delta's$ because the bound state eigenfunctions arising in this situation will play an important r$\hat{\rm o}$le as ground states in the SUSY LLY Hamiltonian, which always encompasses repulsive and attractive $\delta$-point interactions. Section \S.3 is devoted to the generic formulation of the supersymmetric LLY system. In Section \S.4
the specific superpotential defining the supersymmetric LLY Hamiltonian is chosen. Collision states as well as bound states are described
for $N=2$, $N=3$, and a undetermined number of particles. The existence of one ground state of bose type an another one of fermi type will be found, meaning that supersymmetry is unbroken even though the Witten index is zero. Finally, in Section \S.5 some possible continuations
of the problems dealt with in this paper are suggested.

\section{Quantum dynamics of $N$ Lieb-Liniger-Yang bosonic particles on a line}

Consider a system of $N$ bosons that move along the real axis and whose dynamics is determined by the Hamiltonian \cite{Yang,Takahashi,Ushveridze}
\begin{equation}
\hat{H}_N=-\sum_{j=1}^N\frac{\partial^2}{\partial x_j^2}+2c\times\sum_{1\leq j<k \leq N}\delta(x_j-x_k) \ , \label{hamYangN}
\end{equation}
where $x_j$ represents the coordinate of the $j$-th particle.
\\
Interactions among particles are thus considered as ultra-short range -contact interactions- and are modeled in terms of Dirac delta functions, namely $2\, c \cdot \,\delta(x_j-x_k)$. The real parameter $c$ characterizes the strength of interactions, $c = 0$ is associated to free particles, $c < 0$ indicates attraction between particles when colliding whereas $c > 0$ represents repulsive collisions. Particles can pass through each other except at the impenetrable limit $c=\infty$.

\subsection{$N$ bosons: the Lieb-Liniger-Yang system of $N$ particles}

The Lieb-Liniger-Yang's Hamiltonian \eqref{hamYangN} gather the interactions between $N$-bosons moving on the real line by using Dirac delta potentials that respond to the possible $\binom{N}{2}=\frac{N(N-1)}{2}$ contacts between pairs of $N$-particles. The configuration space $(x_1,x_2, \ldots,x_N)\in\mathbb{R^N}$ is divided into $N!$ regions $R_Q: x_{Q_1}<x_{Q_2}<\cdots<x_{Q_N}$ according to the various orders taken by the $N$ bosons over the real line, considering ascending order from left to right.
\\
$Q=(Q_1,Q_2,\ldots,Q_N)\in S_N$ represents one of the permutations of  $(1,2,\ldots,N)$  whereas $S_N$ is the symmetric group of order $N!$. Two regions $R_Q$ and $R_{Q\prime}$ are contiguous, sharing a common boundary on the $N-1$-dimensional hyper-plane, namely $x_{Q_i}=x_{Q_{i+1}}$ if $x_{Q_i}=x_{Q^\prime_{i+1}}$ and $x_{Q_{i+1}}=x_{Q^\prime_{i}}$: there has been an exchange between particles in $x_{Q_i}$ and $x_{Q_{i+1}}$ to go from $R_Q$ to $R_{Q^\prime}$.
We consider here the region $R_I$ as the starting point. $R_I$ is characterized by the identity permutation $I=(1,2,\ldots,N)$ with natural order $x_1<x_2<\cdots<x_N$.

\subsubsection{Bethe wave functions of $N$ bosons: collision states}

We investigate the existence of states characterized by $N$ momenta, which are organized in a decreasing order from left to right: $k_1>k_2>\cdots>k_N$. Bethe ansatz suggests the following wave function over the region $R_Q$ which is symmetric with respect to any boson exchange:
\begin{equation*}
\psi_Q(x_1,x_2,\ldots,x_N)=\sum_P\alpha(P_1P_2 \cdots P_N)e^{i(k_{P_1}x_{Q_1}+k_{P_2}x_{Q_2}+\cdots+k_{P_N}x_{Q_N})} \ . \label{wfbnp}
\end{equation*}
The summation covers all permutations $P=(P_1,P_2,\ldots,P_N)\in S_N$ of the momenta $(k_1,k_2,\ldots,k_N)$ and the coefficients of the linear combination are specified by imposing the matching conditions at boundaries for each permutation $P$. In the complete configuration space the wave function $\psi(x_1,x_2,\ldots,x_N)$ is obtained from the wave functions at every single region by using the appropriate step functions
\begin{equation*}
\psi(x_1,x_2,\ldots,x_N)=\sum_{Q}\theta(x_{Q_2}-x_{Q_1})\theta(x_{Q_3}-x_{Q_2})\cdots \theta(x_{Q_N}-x_{Q_{N-1}})\psi_Q(x_1,x_2,\ldots,x_N) \ . \label{wfgen}
\end{equation*}
At this point, the summation covers the permutations of the $N$ positions $x_i$, $i=1,2, \ldots, N$.
\\
Let us now consider two regions $R_Q$ and $R_{Q^\prime}$ such that $x_{Q_i}=x_{Q^\prime _{i+1}}=x_j$ y $x_{Q_{i+1}}=x_{Q^\prime_ i}=x_k$ for some $i=1,\ldots,N-1$, i.e., two regions, related to each other by two-particle exchange, that share a common boundary, namely the $N-1$-dimensional hyper-plane $x_j=x_k$. In this hyper-plane the only non-zero interaction term in the Hamiltonian $\hat{H}_N$ is $2c\delta(x_j-x_k)$, contact is only possible between particles located at $x_j$ and $x_k$. It is then necessary to study the effective spectral problem \cite{Hallnas}
\begin{equation*}
\left[-\sum_{l=1}^{i-1}\frac{\partial^2}{\partial x_l^2}-\frac{\partial^2}{\partial x_j^2}-\frac{\partial^2}{\partial x_k^2}-\sum_{l=i+2}^N\frac{\partial^2}{\partial x_l^2}+2c\delta(x_j-x_k)\right]\psi(x_1,x_2,\ldots,x_N)=E\psi(x_1,x_2,\ldots,x_N) \ .
\end{equation*}
Over the region $R_Q$, where $x_j<x_k$, the wave function is of the form
\begin{align*}
\psi_Q(x_1,x_2,\ldots,x_N)=&\sum_P\, \alpha(P_1 P_2 \cdots P_iP_{i+1} \cdots P_N)\times \nonumber\\
&\times{\rm exp}[ik_{P_1}x_{Q_1}+ik_{P_2}x_{Q_2}+\cdots +ik_{P_i}x_j+ik_{P_{i+1}}x_k+\cdots  +ik_{P_N}x_{Q_N}] \ ,
\end{align*}
similarly, over $R_{Q^\prime}$, $x_k<x_j$ the wave function is
\begin{align*}
\psi_{Q^\prime}(x_1,x_2,\ldots,x_N)=&\sum_P\, \alpha(P_1 P_2 \cdots P_{i+1}P_{i} \cdots P_N)\times \nonumber\\
&\times{\rm exp}[ik_{P_1}x_{Q^\prime_1}+ik_{P_2}x_{Q^\prime_2}+\cdots +ik_{P_{i+1}}x_j+ik_{P_{i}}x_k+\cdots + ik_{P_N}x_{Q^\prime_N}] \ .
\end{align*}
The exchange or collision of the two particles yields an eigenfunction of the Lieb-Liniger-Yang's Hamiltonian  if the maching conditions
\begin{subequations}
\begin{equation}
\left.\psi\right|_{x_j-x_k\rightarrow0^+}= \left.\psi\right|_{x_j-x_k\rightarrow0^-} \ ,
\end{equation}
\begin{equation}
\left(\frac{\partial}{\partial x_j}-\frac{\partial}{\partial x_k}\right)\left.\psi\right|_{x_j-x_k\rightarrow0^+}-
\left(\frac{\partial}{\partial x_j}-\frac{\partial}{\partial x_k}\right)\left.\psi\right|_{x_j-x_k\rightarrow0^-}=2c\left.\psi\right|_{x_j=x_k} \ ,
\end{equation} \label{discontN}
\end{subequations}
are satisfied.
\\
The inter-coefficient relations that guarantee the fulfilment of (\ref{discontN}) are given by:
\begin{equation}
\frac{\alpha(P_1\cdots P_{i+1}P_i\cdots P_N)}{\alpha(P_1 \cdots P_iP_{i+1} \cdots P_N)}=\frac{i\left(k_{P_{i+1}}-k_{P_i}\right)-c}{i\left(k_{P_{i+1}}-k_{P_i}\right)+c} \ . \label{recN}
\end{equation}
There is a number equal to $N!(N-1)/2$ of independent relations of this type for the $N!$ coefficients $\alpha_P$. This number results from counting all the pairs of permutations that are different after the transposition of two consecutive elements. The factor $N!$ counts the total number of particle position permutations, whereas the factor $N-1$ represents the number of transpositions between consecutive elements within a given permutation. Finally, a factor $1/2$ is included because the relations between coefficients from the transposition, $i$ to $i+1$ or $i+1$ to $i$ are inverse from each other.
\\
The elements of the scattering matrix by definition relate two consecutive coefficients \cite{Ladislav}
\begin{equation*}
\alpha(P_1\cdots P_{i+1}P_i\cdots P_N)=S\left[k_{P_i},k_{P_{i+1}}\right]\alpha(P_1 \cdots P_iP_{i+1}\cdots P_N) \ .
\end{equation*}
Henceforth, from \eqref{recN}, one easily generalizes the two-body S-matrix (\ref{msdpy1}) to $N$-bosons:
\begin{equation*}
S\left[k_{P_i},k_{P_{i+1}}\right]=\frac{i\left[k_{P_{i+1}}-k_{P_i}\right]-c}{i\left[k_{P_{i+1}}-k_{P_i}\right]+c}=e^{i\theta\left(k_{P_{i+1}}
-k_{P_i}\right)} \ ,
\end{equation*}
\begin{equation*}
\theta\left(k_{P_{i+1}}-k_{P_i}\right)=\pi -2{\rm arctan}\frac{k_{P_{i+1}}-k_{P_i}}{c}=-\theta\left(k_{P_{i}}-k_{P_{i+1}}\right).
\end{equation*}
The formulas \eqref{recN} also permit identifying all coefficients. If $j<l$,
\[
\alpha(P_1 P_2 \cdots P_N)={\rm exp}\left[-\frac{i}{2}\sum_{j=1}^N\sum_{l=1}^N \theta (k_{P_l}-k_{P_j})\right]=\prod_{j<l}\, \left[\frac{i(k_{P_l}-k_{P_j})-c}{i(k_{P_l}-k_{P_j})+c}\right].
\]
It is clear that the solution of Lieb-Liniger-Yang's $N$-boson system following the Bethe ansatz, extends the very important property of integrability from $2$ to $N$ particles. Moreover, the LLY $S$-matrix exhibits the even more important property of \underline{factorization} derived from the existence of two equivalent ways of three-particle exchange:
\[
\alpha_{P_1P_2\cdots P_N} \, \hspace{0.3cm} \longrightarrow \, \hspace{0.3cm} \left\{\begin{array}{c} \alpha_{P_1 P_2\cdots P_{i+1}P_iP_{i+2} \cdots P_N} \longrightarrow \alpha_{P_1 P_2\cdots P_{i+1}P_{i+2}P_1 \cdots P_N} \\ \alpha_{P_1P_2\cdots P_iP_{i+2}P_{i+1}\cdots P_N} \longrightarrow \alpha_{P_1P_2\cdots P_{i+2}P_iP_{i+1}\cdots P_N} \end{array}\right\} \, \hspace{0.3cm} \longrightarrow \, \hspace{0.3cm} \alpha_{P_1 P_2\cdots P_{i+2}P_{i+1}P_i \cdots P_N} \]
Thus, the elements of $S$ matrix satisfy the tautological identity
\begin{equation}
S(k_{P_i},k_{P_{i+1}})S(k_{P_i},k_{P_{i+2}})S(k_{P_{i+1}}, k_{P_{i+2}})=S(k_{P_{i+1}},k_{P_{i+2}})S(k_{P_i},k_{P_{i+2}})S(k_{P_i}, k_{P_{i+1}}) \ . \label{fact}
\end{equation}
Of course, factorization requires at least three particles and it is higly non-trivial if the $S(k_i,k_j)$ elements become non-commutative
matrices themselves. In this context the factorization (\ref{fact}) can be seen as the embryo of the Yang-Baxter's algebra that governs the behavior of a Lieb-Liniger-Yang's $N$-particle systems with internal degrees of freedom.
\\
Regarding integrability in the system, the general solution is set as a function of the $N$ quantum numbers $k_1, k_2, \ldots, k_N$, with no room for diffractive processes.
Then, there are $N$ independent symmetry operators that can be chosen, such as
\[
\hat{I}_n=\sum_{j=1}^N\, \hat{p}_j^n =\sum_{j=1}^N\, \left(-i\hbar\frac{\partial}{\partial x_j}\right)^n \quad , \quad n=1,2, \ldots, N \ .
\]

\subsection{$N=2$ bosons: the Lieb-Liniger-Yang system of two bosonic particles}

The dynamics of two bosonic particles, with no spin, moving along the real axis is governed by the Hamiltonian
\begin{equation*}
\hat{H}_2=-\frac{\partial^2}{\partial x_1^2}-\frac{\partial^2}{\partial x_2^2}+2c\cdot\delta(x_1-x_2) \quad , \quad c\in\mathbb{R} \ .\label{hamYang2}
\end{equation*}
\\
The configuration space, herein the space for two points $(x_1,x_2)\in\mathbb{R}^2$ over the real axis, is divided into two regions according to the inequalities:
\begin{equation*}
R_{12}: \ x_1<x_2 \quad , \quad R_{21}: \ x_2<x_1 \, \, ,
\end{equation*}
depending on whether the first point is to the right or to the left of the second point. The boundary between the two regions is the line $x_1=x_2$.

\subsubsection{Collision states}

Since function $\delta(x_1-x_2)$ cancels out in regions $R_{12}$ and $R_{21}$, the eigen-value problem
\begin{equation}
\left[-\frac{\partial^2}{\partial x_1^2}-\frac{\partial^2}{\partial x_2^2}+2c\delta(x_1-x_2)\right]\psi(x_1,x_2)=E\psi(x_1,x_2) \ ,  \label{evp2p}
\end{equation}
is equivalent to having a Helmholtz equation for free particles
\begin{equation*}
\left[-\frac{\partial^2}{\partial x_1^2}-\frac{\partial^2}{\partial x_2^2}\right]\psi(x_1,x_2)=E\psi(x_1,x_2) \quad , \quad x_1\neq x_2 \ , \label{ech}
\end{equation*}
over these two regions. However, at the boundary line $x_1=x_2$ between $R_{12}$ and $R_{21}$, there is an infinite potential. The eigenfunctions must be properly matched through this frontier between the two regions. It is clear that the contact interaction modelled with the Dirac delta potential in (\ref{evp2p}) requires continuity of the wave function and discontinuity of its derivative at the boundary line between regions $R_{12}$ and $R_{21}$. Thus, the spectral problem is globally defined through the matching conditions at $x_1=x_2$:
\begin{subequations}
\begin{equation}
\left.\psi\right|_{x_1-x_2\rightarrow0^+}= \left.\psi\right|_{x_1-x_2\rightarrow0^-} \ ,
\end{equation}
\begin{equation}
\left(\frac{\partial\psi}{\partial x_1}-\frac{\partial\psi}{\partial x_2}\right)\bigg|_{x_1-x_2\rightarrow0^+}-
\left(\frac{\partial\psi}{\partial x_
1}-\frac{\partial\psi}{\partial x_2}\right)\bigg|_{x_1-x_2\rightarrow0^-}=2c\left.\psi\right|_{x_1=x_2} \ .
\end{equation}\label{discont2}
\end{subequations}
\\
By using the Heaviside step function\footnote{The Heaviside step function $H(x)$ is defined as:
$
H(x)=\left\{ \begin{array}{ll}
1 & , \ \ x>0 \\
0 & , \ \ x<0 \\
\end{array} \right. \label{heaviside} \, \, .
$}
$H(x)$, a general wave function for the two-particle system can be written as
\begin{equation}
\psi(x_1,x_2)=H(x_2-x_1)\psi_{12}(x_1,x_2)+H(x_1-x_2)\psi_{21}(x_1,x_2) \ , \label{abdpfg}
\end{equation}
where $\psi_{12}(x_1,x_2)$ and $\psi_{21}(x_1,x_2)$ are the wave functions over regions $R_{12}$ y $R_{21}$ respectively.
\\
Bosonic statistics requires that the wave function be symmetric with respect to two-particle exchange, namely: $\psi(x_1,x_2)=\psi(x_2,x_1)$. The Bethe ansatz for bosons selects the wave functions on regions $R_{12}$ and $R_{21}$ as the linear combinations:
\begin{equation}
\psi_{12}(x_1,x_2)=\alpha_{12}e^{i(k_1 x_1+k_2 x_2)}+\alpha_{21}e^{i(k_2 x_1+k_1 x_2)} \ , \label{abdpf}
\end{equation}
\begin{equation}
\psi_{21}(x_1,x_2)=\alpha_{12}e^{i(k_2 x_1+k_1 x_2)}+\alpha_{21}e^{i(k_1 x_1+k_2 x_2)} \ , \label{abdpq}
\end{equation}
because plane waves with dispersion relation $E=k_1^2+k_2^2$ solve the Helmoltz equation in both $R_{12}$ and $R_{21}$ and (\ref{abdpf})-(\ref{abdpq}) comply with Bose statistics: $\psi_{12}(x_2,x_1)=\psi_{21}(x_1,x_2)$, $\psi_{21}(x_2,x_1)=\psi_{12}(x_1,x_2)$. Note that the $x_1 \leftrightarrow x_2$ exchange also exchanges $R_{12}$ and $R_{21}$.   Since $\psi_k(x=0^+)=\psi_k(x=0^-)=\alpha_{12}+\alpha_{21}$ and $\psi^\prime_k(x=0^+)-\psi^\prime_k(x=0^-)=ik(\alpha_{12}-\alpha_{21})$, the matching conditions are satisfied if and only if
\begin{equation}
\frac{\alpha_{21}}{\alpha_{12}}=\frac{ik-c}{ik+c}=\frac{i(k_2-k_1)-c}{i(k_2-k_1)+c} \label{betec} \ .
\end{equation}
It is clear that the quotient of the coefficients in (\ref{betec}) is a complex unit-modulus number, therefore
\[
\frac{\alpha_{21}}{\alpha_{12}}=e^{i\theta(k_2-k_1)} \quad , \quad \theta(k)=\pi-2{\rm arctan}\frac{k}{c}\in(0,2\pi)
\]
where the arctan function has been limited to its principal defining interval, namely $(-\frac{\pi}{2},\frac{\pi}{2})$.
\\
It is convenient to write the coefficients $\alpha_{12}$ and $\alpha_{21}$ in the form
\begin{equation*}
\alpha_{12}=e^{\frac{i}{2}\theta(k_1-k_2)} \quad , \quad \alpha_{21}=e^{\frac{i}{2}\theta(k_2-k_1)} \ . \label{coef}
\end{equation*}
Bethe wave functions over regions $R_{12}$ and $R_{21}$ become:
\begin{equation}
\psi_{12}(x_1,x_2)=e^{\frac{i}{2}\theta(k_1-k_2)}\left[e^{i(k_1 x_1+k_2 x_2)}+e^{i(k_2 x_1+k_1 x_2)+i\theta(k_2-k_1)}\right]  \ ,\label{mon1}
\end{equation}
\begin{equation}
\psi_{21}(x_1,x_2)=e^{\frac{i}{2}\theta(k_1-k_2)}\left[e^{i(k_2 x_1+k_1 x_2)}+e^{i(k_1 x_1+k_2 x_2)+i\theta(k_2-k_1)}\right] \ . \label{mon2}
\end{equation}
The solution over the region $R_{12}$, $x_1<x_2$, is a linear combination of two terms: (1) $e^{i(k_1 x_1+k_2 x_2)}$ where the two particles propagate as plane waves with momenta $k_1$ and $k_2$ and (2) $e^{i(k_2 x_1+k_1 x_2)+i\theta(k_2-k_1)}$ where the momenta are now exchanged according to Bose's statistics. A relative phase shift results from this exchange. Over the region $R_{21}$, $x_1>x_2$, there is a similar solution with exchanged momenta and the matching conditions at the boundary dictate the exchange phase. The reduction of the problem to mass center allows a new interpretation of this exchange interaction as an induced phase shift caused by the Dirac delta potential $\delta(x_1-x_2)$  that takes over when one of the particles surpasses the other particle across the boundary between $R_{12}$ and $R_{21}$ \cite{Ladislav}. In this process, the total momentum $K=k_1+k_2$ and the total energy $E=k_1^2+k_2^2$ are conserved quantities, thus there is elastic scattering so that particles exchange their momenta only.  If the momenta over $R_{12}$ are initially $k_1$ and $k_2$, the initial momenta over $R_{21}$ are $k_1^\prime=k_2$ and $k_2^\prime=k_1$ \cite{Ladislav}. In the literature, the exchange phase is commonly written as a scattering matrix although it can be identified as a phase shift in the even channel caused by the potential of the reduced problem{\footnote{It can be stated that it is the determinant of the $S$ matrix, since for a delta potential the phase shift in the odd channel is zero, therefore the other eigenvalue equals 1.
}}.
\begin{equation*}
\alpha_{21}(k_2,k_1)=S(k_1,k_2)\alpha_{12}(k_1,k_2)  \quad , \quad
S(k_1,k_2)=\frac{i(k_2-k_1)-c}{i(k_2-k_1)+c}=e^{i\theta(k_2-k_1)} \ . \label{msdpy1}
\end{equation*}

\subsubsection{Bound states}

When $c<0$ the relative problem also admits a bound state and our goal in this sub-Section is to analyze the bound state wave functions that solve the two-body problem when the $\delta$-interaction is attractive..

T
\\
In summary, there are three types of eigenstates associated to the Hamiltonian of the two-particle Lieb-Liniger-Yang´s system when $c<0$:

\underline{Ground state}: the ground state is the bound state of the two particles with no movement of their mass center: $P=0$.
The ground state energy, which is the smallest eigenvalue of the Hamiltonian operator, and the ground state wave function are:
\begin{equation*}
E_0=-\frac{|c|^2}{2} \, \, \, \, , \, \, \, \quad \quad \quad
\psi_0(x)=2|c|e^{-\frac{|c|}{2}|x|} \quad , \quad c<0 \ .   \label{esf}
\end{equation*}
\underline{Dimer-type excited states}: as stated above, the only difference compared to the ground state is the movement of the mass center, with momentum $K=2P\neq 0$ so that the energy is
\begin{equation*}
E=E_0+2P^2 \ . \label{eexl2}
\end{equation*}
\underline{Monomer-type excited states}: these states also come from the scattering solutions to the reduced problem. Their energy is always positive {\footnote{A possible bound state exactly lying at the continuous spectrum threshold, where energy should be zero, is excluded here.}}
\begin{equation*}
E=\frac{K^2}{2}+\frac{k^2}{2} >0 \quad , \quad k\in\mathbb{R}   \label{eexs2}
\end{equation*}

Wave functions of this type lead to an amplitude with non-zero probability to find the two particles infinitely separated from one another, see formulas (\ref{mon1}) and (\ref{mon2}). This guarantees the individual nature of each particle and also justifies the use of the term monomer to refer to each entity. The exchange phases reflect the collisions between two monomers.
\\
If $c>0$, only the third type remains, while the ground state comes from the continuous spectrum threshold found in the two-body relative problem.

\subsection{$N=3$ bosons: the Lieb-Liniger-Yang system of three bosonic particles}

The dynamics of three bosonic particles, with no spin, moving along the real axis is governed by the Hamiltonian
\begin{equation}
\hat{H}=-\bigtriangleup+2c\delta(x_1-x_2)+2c\delta(x_1-x_3)+2c\delta(x_2-x_3) \ , \label{hamYang3}
\end{equation}
where $c\in\mathbb{R}$ but now $\bigtriangleup=\sum_{k=1}^{N=3}\frac{\partial^2}{\partial x_k^2}$ is the Laplacian in $\mathbb{R}^3$ that is the new configuration space.
\\
The configuration space, herein the space for three points $(x_1,x_2,x_3)\in\mathbb{R}^3$ over the real axis, is divided into six regions $R_Q$ according to the inequalities:
\begin{equation*}
\begin{array}{ccc}
R_{123}: \ x_1<x_2<x_3 \quad , \quad R_{132}: \ x_1<x_3<x_2 \quad , \quad R_{213}: \ x_2<x_1<x_3\\
\ \\
R_{231}: \ x_2<x_3<x_1 \quad , \quad R_{312}: \ x_3<x_1<x_2 \quad , \quad R_{321}: \ x_3<x_2<x_1 \label{ras}
\end{array}
\end{equation*}
associated to six permutations of three elements:
\[
Q=\left(\begin{array}{ccc} 1 & 2 & 3 \\ Q_1 & Q_2 & Q_3 \end{array}\right) .
\]

\subsubsection{Collision states}
The spectral problem 
\begin{equation*}
\left[-\bigtriangleup+2c\delta(x_1-x_2)+2c\delta(x_1-x_3)+2c\delta(x_2-x_3)\right]\psi(x_1,x_2,x_3)=E\psi(x_1,x_2,x_3) \ . \label{evp3}
\end{equation*}
is solved using the Bethe ansatz in each region $R_{Q}$:
\begin{align}
\psi_{Q_1 Q_2 Q_3}(x_1,x_2,x_3)=& \ \alpha_{123}e^{i(k_1 x_{Q_1}+k_2 x_{Q_2}+k_3 x_{Q_3})}+\alpha_{213}e^{i(k_2 x_{Q_1}+k_1 x_{Q_2}+k_3 x_{Q_3})}+\nonumber\\
&+\alpha_{132}e^{i(k_1 x_{Q_1}+k_3 x_{Q_2}+k_2 x_{Q_3})}+\alpha_{321}e^{i(k_3 x_{Q_1}+k_2 x_{Q_2}+k_1 x_{Q_3})}+\nonumber\\
&+\alpha_{312}e^{i(k_3 x_{Q_1}+k_1 x_{Q_2}+k_2 x_{Q_3})}+\alpha_{231}e^{i(k_2 x_{Q_1}+k_3 x_{Q_2}+k_1 x_{Q_3})} \ . \label{abtp} 
\end{align}
The eigenfunctions are
\begin{align*}
\psi(x_1,x_2,x_3)=&\hspace{0.2cm}\theta(x_2-x_1)\theta(x_3-x_2)\psi_{123}(x_1,x_2,x_3)+\theta(x_1-x_2)\theta(x_3-x_1)\psi_{213}(x_1,x_2,x_3)+ \nonumber\\
&+\theta(x_3-x_1)\theta(x_2-x_3)\psi_{132}(x_1,x_2,x_3)+\theta(x_2-x_3)\theta(x_1-x_2)\psi_{321}(x_1,x_2,x_3)+ \nonumber\\
&+\theta(x_1-x_3)\theta(x_2-x_1)\psi_{312}(x_1,x_2,x_3)+\theta(x_3-x_2)\theta(x_1-x_3)\psi_{231}(x_1,x_2,x_3) \ , \label{abtpg} 
\end{align*}
while the coefficients $\alpha_{P_1P_2P_3}(k_1,k_2,k_3)$ are determined using matching conditions, i.e. continuity of wave functions and discontinuity of its derivative at boundary line between two regions $R_{Q}$. 
\\
For example, in the regions $R_{123}$ and $R_{213}$ whose frontier is the plane $x_1=x_2$, the effective interaction in the Hamiltonian \eqref{hamYang3} is $2c\delta(x_1-x_2)$ and so it is necessary to study the spectral problem constrained to $R_{123}\bigcup R_{213}$ . In this case, the effective Hamiltonian is:
\begin{equation*}
\left[-\sum_{j=1}^3\frac{\partial^2}{\partial x_j^2}+2c\delta(x_1-x_2)\right]\psi(x_1,x_2,x_3)=E\psi(x_1,x_2,x_3) \ . \label{espef3}
\end{equation*}
The eigenfunctions of this simplified spectral problem are
\begin{equation*}
\psi(x_1,x_2,x_3)=\theta(x_2-x_1)\psi_{123}(x_1,x_2,x_3)+\theta(x_1-x_2)\psi_{213}(x_1,x_2,x_3) \ , \label{wfgen3}
\end{equation*}
where  the wave functions  in $R_{123}$ and $R_{213}$ are of the form
\begin{equation}
\psi_{123}(x_1,x_2,x_3)=\phi_{123}(x_1,x_2)e^{i(k_3x_3)}=\left[\alpha_{123}e^{i(k_1x_1+k_2x_2)}+\alpha_{213}e^{i(k_2x_1+k_1x_2)}\right]e^{ik_3x_3} \ , \label{ab2wf}
\end{equation}
\begin{equation}
\psi_{213}(x_1,x_2,x_3)=\phi_{213}(x_1,x_2)e^{i(k_3x_3)}=\left[\alpha_{123}e^{i(k_1x_2+k_2x_1)}+\alpha_{213}e^{i(k_2x_2+k_1x_1)}\right]e^{ik_3x_3} \ , \label{ab2wf2}
\end{equation}
and the functions $\phi_{123}(x_1,x_2)$ and $\phi_{213}(x_1,x_2)$ are eigenfunctions of the Hamiltonian
\begin{equation*}
\hat{H}=-\frac{\partial^2}{\partial x_1^2}-\frac{\partial^2}{\partial x_1^2}+2c\delta(x_1-x_2) \ .
\end{equation*}
The matching conditions are satisfied if and only if
\begin{equation}
\frac{\alpha_{213}}{\alpha_{123}}=\frac{i(k_2-k_1)-c}{i(k_2-k_1)+c}=e^{i\theta(k_2-k_1)} \  \label{alphap}
\end{equation}
In the other regions $R_Q$ and $R_P$ the procedure is similar, the resuls are:
\begin{equation}
\frac{\alpha_{213}}{\alpha_{123}}=\frac{\alpha_{321}}{\alpha_{312}}=\frac{i(k_2-k_1)-c}{i(k_2-k_1)+c}=e^{i\theta(k_2-k_1)}=S(k_1,k_2) \ ,  \label{alphaq1}
\end{equation}
\begin{equation}
\frac{\alpha_{312}}{\alpha_{132}}=\frac{\alpha_{231}}{\alpha_{213}}=\frac{i(k_3-k_1)-c}{i(k_3-k_1)+c}=e^{i\theta(k_3-k_1)}=S(k_1,k_3) \ , \label{alphaq2}
\end{equation}
\begin{equation}
\frac{\alpha_{321}}{\alpha_{231}}=\frac{\alpha_{132}}{\alpha_{123}}=\frac{i(k_3-k_2)-c}{i(k_3-k_2)+c}=e^{i\theta(k_3-k_2)}=S(k_2,k_3) \ . \label{alphaq3}
\end{equation} 
This relations can be written in the form
\[
\alpha_{P_1P_2P_3}=\frac{\left[\hspace{0.05cm}i(k_{P_2}-k_{P_1})-c\hspace{0.06cm}\right]}{\left[\hspace{0.05cm}i(k_{P_2}-k_{P_1})+c\hspace{0.06cm}\right]}
\cdot\frac{\left[\hspace{0.05cm}i(k_{P_3}-k_{P_1})-c\hspace{0.06cm}\right]}{\left[\hspace{0.05cm}i(k_{P_3}-k_{P_1})+c\hspace{0.06cm}\right]}\cdot
\frac{\left[\hspace{0.05cm}i(k_{P_3}-k_{P_2})-c\hspace{0.06cm}\right]}{\left[\hspace{0.05cm}i(k_{P_3}-k_{P_2})+c\hspace{0.06cm}\right]}
\]
where $(P_1,P_2,P_3)$ is a permutation of  the momenta.
\\
Bethe wave functions over regions $R_{123}$ and $R_{213}$ become:
\begin{equation*}
\psi_{123}(x_1,x_2,x_3)=e^{\frac{i}{2}[\theta(k_1-k_2)+\theta(k_1-k_3)+\theta(k_2-k_3)]}\left[e^{i(k_1 x_1+k_2 x_2)}+e^{i(k_1 x_2+k_2 x_1)+i\theta(k_2-k_1)}\right]e^{ik_3x_3}  
\end{equation*}
\begin{equation*}
\psi_{213}(x_1,x_2,x_3)=e^{\frac{i}{2}[\theta(k_1-k_2)+\theta(k_1-k_3)+\theta(k_2-k_3)]}\left[e^{i(k_1 x_2+k_2 x_1)}+e^{i(k_1 x_1+k_2 x_2)+i\theta(k_2-k_1)}\right]e^{ik_3x_3}  
\end{equation*}
\\
The solution on the region $R_{123}$, $x_1<x_2<x_3$, corresponds to the product of a plane wave $e^{ik_3x_3}$ -describing the independent movement of the particle at $x_3$- and a linear combination of two terms: (1) $e^{i(k_1 x_1+k_2 x_2)}$ where the particles at  $x_1$ and $x_2$ propagate as plane waves with momenta $k_1$ and $k_2$, and (2)  $e^{i(k_1 x_2+k_2 x_1)+i\theta(k_2-k_1)}$, where now the momenta have been exchanged according to Bose statistics. A relative phase then emerges as a result of the exchange.
On the region $R_{213}$, $x_2<x_1<x_3$, the solution is similar, with exchanged momenta $k_1$ and $k_2$, the matching conditions at the boundary between the two regions govern the exchange phase. In this case, the reduction of the collision problem for particles at $x_1$ and $x_2$ with respect to the mass center allows a reinterpretation of the exchange interaction as the phase shift induced by the Dirac delta potential $\delta(x_1-x_2)$ that applies when  the particle at $x_1$ surpasses the other particle, at $x_2$ across the boundary between $R_{123}$ and $R_{213}$ \cite{Ladislav}. During this process, the total momentum and energy are conserved quantities, thus there is elastic scattering such that the particles at  $x_1$ and $x_2$ only exchange their momenta. If the initial momenta on $R_{123}$ are $k_1$, $k_2$ and $k_3$, then the initial momenta on $R_{213}$ are $k_1^\prime=k_2$, $k_2^\prime=k_1$ and $k_3^\prime=k_3$ \cite{Ladislav}. The solution on other regions, $R_Q$ and $R_P$, where the other two pairs of particles are the ones leading the exchange, is similar.

\subsubsection{Bound states}
\underline{Ground state}: the ground state, symmetric in the permutations of three particles, is the bound state of the three particles with no movement of their mass center: $P=0$, and with energy $E=-2\vert c \vert^2$ 
\begin{equation}
\psi_0(x_1,x_2,x_3)=Ae^{-\frac{|c|}{2}(|x_1-x_2|+|x_1-x_3|+|x_2-x_3|)} \quad , \quad c<0 \label{wf03p}
\end{equation}
where $A$ is a normalization constant.
\\
\underline{Trimer-type excited states}: the only difference compared to the ground state is the movement of the mass center, with momentum $K=3P\neq 0$ so that the energy is $E=E_0+3P^2=-2\vert c \vert^2+3P^2$. The wave functions of these states are
\begin{equation}
\psi_{\rm P}(x_1,x_2,x_3)=e^{3iPX}e^{-\frac{|c|}{2}(|x_1-x_2|+|x_1-x_3|+|x_2-x_3|)} \quad , \quad c<0 
\end{equation}
where $X=\frac{x_1+x_2+x_3}{3}$ is the center of mass coordinate.

\underline{Monomer-dimer-type excited states}: bound state involving only two particles, the third particle, the monomer, moves freely and collides with the dimer. The wave functions of these states are
\begin{eqnarray*}
&& \psi_{\rm PQ}(x_1,x_2,x_3)=\\ &=& {\rm exp}[2i P X_{12}]{\rm exp}\left[-\frac{\vert c\vert}{2}\vert x_1-x_2\vert\right]{\rm exp}[iQ x_3] + 
{\rm exp}[2i P X_{13}]{\rm exp}\left[-\frac{\vert c\vert}{2}\vert x_1-x_3\vert\right]{\rm exp}[iQ x_2]\\&+&
{\rm exp}[2i P X_{23}]{\rm exp}\left[-\frac{\vert c\vert}{2}\vert x_2-x_3\vert\right]{\rm exp}[iQ x_1] \label{wfdm3p}
\end{eqnarray*}
where
\[
X_{12}=\frac{x_1+x_2}{2} \quad , \quad X_{13}=\frac{x_1+x_3}{2} \quad , \quad X_{23}=\frac{x_2+x_3}{2}
\]
are the center of mass coordinates of each air of particles. The momenta and energy are:
\begin{equation}
K_{{\rm m/d}}=k_1+k_2+k_3=Q+2P \quad , \quad E_{\rm m/d}=k_1^2+k_2^2+k_3^2=Q^2+2P^2-\frac{|c|^2}{2}  \ .  \label{ke12}
\end{equation}
\\
\underline{Monomer-type excited states}: these states are formed by three monomers and characterized by three real moments, there are all kinds of collisions or exchanges between the particles and they always have positive energy. 


\section{$\mathcal{N}=2$ extended supersymmetric quantum mechanics: Systems with $N$ degrees of freedom}
The building blocks of supersymmetric quantum mechanics are the supercharges $\hat{Q}_{i}$, a set of $\mathcal{N}$
quantum operators obeying the Heisenberg superalgebra, see e.g. \cite{Freund}:
$$
\hat{Q}_{i}\hat{Q}_{j}+\hat{Q}_{j}\hat{Q}_{i}=\{\hat{Q}_{i},\hat{Q}_{j}\}=2\delta_{ij}\hat{H}, \ \ \ i,j=1,\ldots,\mathcal{N},
$$
$$
\hat{H}\hat{Q}_{i}-\hat{Q}_{i}\hat{H}=[\hat{H},\hat{Q}_{i}]=0, \ \ \ i=1,\ldots,\mathcal{N}.
$$
$\hat{H}$ is the quantum Hamiltonian governing the dynamics and obviously the supercharges are invariants that generate the \lq\lq super-symmetries\rq\rq of the system. In the $\mathcal{N}=2$ case, our playground in this essay, it is convenient to trade the hermitian supercharges $\hat{Q}_{1}$ and $\hat{Q}_{2}$ by the non-hermitian combinations: $\hat{Q}=\frac{1}{\sqrt{2}}(\hat{Q}_{1}+i\hat{Q}_{2})$, $\hat{Q}^{\dagger}=\frac{1}{\sqrt{2}}(\hat{Q}_{1}-i\hat{Q}_{2})$. The Heisenberg superalgebra looks now, see e.g. \cite{Matgui}, in the simpler form:
\begin{equation}
\{\hat{Q},\hat{Q}^{\dagger}\}=2\hat{H}, \quad \qquad
[\hat{H},\hat{Q}]=0,\ \ \ [\hat{H},\hat{Q}^{\dagger}]=0, \ \ \ \{\hat{Q},\hat{Q}\}=0, \ \ \ \{\hat{Q}^{\dagger},\hat{Q}^{\dagger}\}=0,  \label{algqqd}
\end{equation}
and $\hat{Q}$ and $\hat{Q}^{\dagger}$ show themselves as the generators of the super-symmetries of the system.
\\
 To implement this algebraic structure on a system with $N$ \lq\lq bosonic\rq\rq degrees of freedom $\hat{x}_j$ it is necessary to add $2N$ \lq\lq fermionic\rq\rq degrees of freedom $\hat{\psi}_j$, $\hat{\psi}_j^{\dagger}$ such that $(\hat{\psi}_j)^2=(\hat{\psi}^\dagger_j)^2=0 \, \, , \, \, \forall j=1,\ldots,N$. The physical dimension of the \lq\lq fermionic\rq\rq coordinates is $[\hat{\psi}_k]=M^{-\frac{1}{2}}$ -we denote the physical dimension of the observable ${\mathcal O}$ in the form $[\mathcal{O}]$-.
 The position $\hat{x}_j$ and momentum $\hat{p}_j$ operators satisfy the commutation rules
\begin{equation}
\left[\hat{x}_j,\hat{x}_k\right]=0=\left[\hat{p}_j,\hat{p}_k\right],\ \ \ \left[\hat{x}_j,\hat{p}_k\right]= i\hbar\delta_{jk} \ \ \forall \ \  j,k=1,2,\ldots,N \label{relconmNdim}
\end{equation}
in the canonical quantization procedure. In \lq\lq coordinate\rq\rq representation the (bosonic)
Hilbert space of states of the system is the space $L_2(\mathbb{R}^N)$ of square integrable functions from $\mathbb{R}^N$ to $\mathbb{C}$. The action of the $\hat{x}_j$ and $\hat{p}_j$ operators on the wave function $\psi(x_1,\ldots,x_N)\in L_2(\mathbb{R}^N)$ accordingly is:\footnote{Any other representation of the Heisenberg algebra (\ref{relconmNdim}) is unitarily equivalent. Stone-Von Neumann theorem.}
$$
\hat{x}_j\psi(x_1,\ldots,x_N)=x_j\psi(x_1,\ldots,x_N),\ \ \ j=1,\ldots,N
$$
$$
\hat{p}_j\psi(x_1,\ldots,x_N)=-i\hbar\frac{\partial}{\partial x_j}\psi(x_1,\ldots,x_N),\ \ \ j=1,\ldots,N.
$$
The fermionic operators -describing the fermionic degrees of freedom- satisfy the anti-commutation relations
\begin{equation}
\{\hat{\psi}_j,\hat{\psi}_k\}=0=\{\hat{\psi}_j^\dagger,\hat{\psi}_k^\dagger\},\ \ \ \{\hat{\psi}_j,\hat{\psi}_k^\dagger\}=\frac{1}{m}\delta_{jk} \ \ \forall \ \  j,k=1,2,\ldots,N, \label{relanticonmNdim}
\end{equation}
where $m$ is a parameter of the system with dimensions of mass.
\\It is compulsory at this point to clarify an almost certain notational confusion. In Section \S.2 we referred to \lq\lq bosonic\rq\rq or \lq\lq fermionic\lq\lq statistics as they are characterized in quantum mechanics: in response to particle exchanges in a system of $N$ particles the wave function is respectively symmetric or antisymmetric under any permutation of $N$ bosons or fermions. In the sequel we shall rely on the analogy provided by understanding the canonical quantization rules (\ref{relconmNdim}) or (\ref{relanticonmNdim})  as those arising between a set of $N$ bosonic or fermionic fields in $0$-dimensional QFT. We stress that the terminology established above in Section \S.3, to be used in ${\cal N}=2$ supersymmetric quantum mechanics, is based simply in this analogy. However, we warn that use of the old \S.2 meaning of boson or fermion is not discarded in Sections \$.3 and \S.4., the interpretation should be clear from the context.
\\
The supercharges are defined from the \lq\lq super-potential\rq\rq, a function{\footnote{Dealing with delta function potentials we will need to extend the mathematical nature of the super-potentials to the space of distributions in $\mathbb{R}^N$.}}
$W(x_1,x_2,\ldots,x_N):\mathbb{R}^N\rightarrow\mathbb{R}$ with dimension $[W]=ML^2T^{-1}$ of action, and the fermionic operators in the form:
\begin{equation}
\hat{Q}=i\sum_{j=1}^N\hat{\psi}_j\left(\hbar\frac{\partial}{\partial x_j}+\frac{\partial W}{\partial x_j}\right)=
i\sum_{j=1}^N\hat{\psi}_j\hat{D}_j, \quad , \quad
\hat{Q}^\dagger=i\sum_{j=1}^N\hat{\psi}_j^\dagger\left(\hbar\frac{\partial}{\partial x_j}-\frac{\partial W}{\partial x_j}\right)=i\sum_{j=1}^N\hat{\psi}_j^\dagger\hat{D}_j^\dagger, \label{opqdNdim}
\end{equation}
such that $\hat{Q}$ and $\hat{Q}^{\dagger}$  themselves are fermionic operators.
\\
The supersymmetric Hamiltonian is accordingly the differential operator
\begin{align}
\hat{H}=\frac{1}{2}\{\hat{Q},\hat{Q}^\dagger\}&=-\frac{1}{2m}\sum_{j=1}^N\hat{D}_j\hat{D}_j^\dagger \ \mathbb{\hat{I}}_{2^N}-\hbar\sum_{j=1}^N\sum_{k=1}^N\hat{\psi}_k^\dagger\hat{\psi}_j\frac{\partial^2W}{\partial x_k\partial x_j}\nonumber\\
&=-\frac{1}{2m}\sum_{j=1}^N\hat{D}_j^\dagger\hat{D}_j \ \mathbb{\hat{I}}_{2^N}+\hbar\sum_{j=1}^N\sum_{k=1}^N\hat{\psi}_k\hat{\psi}_j^\dagger\frac{\partial^2W}{\partial x_k\partial x_j}, \label{hamNdim}
\end{align}
where $\mathbb{\hat{I}}_{2^N}$ is the $2^N\times2^N$ identity matrix. The reason for the name super-potential is clear here.
In formula (\ref{hamNdim}) the Yukawa couplings are determined from the second partial derivatives of $W$. The remaining interaction energies are seen in
\[
\sum_{j=1}^N\hat{D}_j^\dagger\hat{D}_j=\hbar^2 \bigtriangleup-\sum_{j=1}^N\partial_jW\partial_jW + \hbar \bigtriangleup W \quad .
\]
\\
 The (fermionic) Hilbert space is the finitely generated fermionic Fock space: $\mathcal{F}=\bigoplus_{j=0}^N\mathcal{F}_j=\mathcal{F}_0\oplus\mathcal{F}_1\oplus\cdots\oplus\mathcal{F}_N$. The structure of $\mathcal{F}$ is as follows:
 \begin{enumerate}
 \item Everything is based on the vacuum state:  $\left|0\right\rangle$ belonging to the kernel of \underline{all}
 the annihilation fermionic operators $\hat{\psi}_j\left|0\right\rangle=0$, $\forall j=1,2,\ldots,N$. The vacuum state also lives in the kernel of the fermionic number operator $\hat{F}|0\rangle=\left[\sum_{k=1}^N \, \hat{\psi}_k^\dagger\hat{\psi}_k\right]|0\rangle =0|0\rangle$. The interpretation is clear: in this state all the fermionic degrees of freedom are unoccupied. $|0\rangle$ is the unique state in the basis of $\mathcal{F}_0$ which is thus one-dimensional.
 \item One-fermion states: $\left|1_j\right\rangle=\hat{\psi}_j^\dagger\left|0\right\rangle$, $j=1,2,\ldots,N$.
 All these $N$-states obtained from the vacuum by the action of one of the creation operators $\hat{\psi}_j^\dagger$ are eigen-states of the fermionic number operator of eigenvalue 1: $\hat{F}|1_k\rangle=|1_k\rangle$. They correspond to states where one of the fermionic degrees of freedom is occupied and form an ortho-normal basis, $\langle 1_j |1_k\rangle =\delta_{jk}$, of $\mathcal{F}_1$ which has dimension $N$.

 \item Two-fermion states: $\left|1_{j_2}1_{j_1}\right\rangle=\hat{\psi}_{j_2}^\dagger\hat{\psi}_{j_1}^\dagger\left|0\right\rangle$, ${j_1},{j_2}=1,2,\ldots,N $ ($j_2>j_1$). Because $\hat{F}\left|1_{j_2}1_{j_1}\right\rangle=2\left|1_{j_2}1_{j_1}\right\rangle$ in these states two fermionic degrees of freedom are occupied. The anti-commutation rules imply anti-symmetry with respect to the order of occupation $|1_{j_2}1_{j_1}\rangle =-|1_{j_1}1_{j_2}\rangle $ and, this, in turn, means that the individual fermionic occupation number $\hat{F}_k=\hat{\psi}_k^\dagger\hat{\psi}_k$ has eigenvalues $0$ and $1$, the exclusion principle. There are $\binom{N}{2}=\frac{N(N-1)}{2}$ states of this kind that form an orthonormal basis in $\mathcal{F}_2$: $ \langle 1_{k_2}1_{k_1}|1_{j_2}1_{j_1}\rangle=\delta_{k_1j_1}\delta_{k_2j_2}$.

 \item  $n$-fermion states: the action of $n$ creation operators over $|0\rangle$ gives rise to $\binom{N}{n}=\frac{N!}{(N-n)!n!}$ $n$-fermion states $\left|1_{j_n}1_{j_{n-1}}\cdots1_{j_1}\right\rangle=
     \hat{\psi}_{j_n}^\dagger\hat{\psi}_{j_{n-1}}^\dagger\cdots\hat{\psi}_{j_1}^\dagger
\left|0\right\rangle$, ${j_1},{j_2}\ldots{j_n}=1,2,\ldots,N$ ($j_n>j_{n-1}>\cdots>j_1$). Because $\hat{F}\left|1_{j_n}1_{j_{n-1}}\cdots1_{j_1}\right\rangle=n\left|1_{j_n}1_{j_{n-1}}\cdots1_{j_1}\right\rangle$ in these states $n$ fermionic degrees of freedom are occupied. The anti-commutation relations force  anti-symmetry  under the exchange of the occupation order of $j_a$ and $j_b$, i.e., $\left|1_{j_n}\cdots1_{j_a}1_{j_b}\cdots1_{j_1}\right\rangle=
-\left|1_{j_n}\cdots1_{j_b}1_{j_a}\cdots1_{j_1}\right\rangle$. The following ortho-normality conditions are satisfied
 $$
\left\langle\left.1_{k_n}1_{k_{n-1}}\cdots1_{k_1}\right|1_{j_n}1_{j_{n-1}}\cdots1_{j_1}\right\rangle=
\delta_{k_nj_n}\delta_{k_{n-1}j_{n-1}}\cdots\delta_{k_1j_1} \quad .
$$
Accordingly, the $\left(\begin{array}{c}N \\ n\end{array}\right)$ $\left|1_{j_n}1_{j_{n-1}}\ldots1_{j_1}\right\rangle$ states form a basis in $\mathcal{F}_n$ and, therefore, an arbitrary state in the sub-space $\mathcal {F}_n$ of the fermionic Fock space is of the form:
$$
\left|F_n\right\rangle=\sum_{j_n>j_{n-1}>\cdots>j_1}^N f_{j_nj_{n-1}\cdots j_1}\left|1_{j_n}1_{j_{n-1}}\cdots1_{j_1}\right\rangle \quad , \quad f_{j_nj_{n-1}\cdots j_1}\in\mathbb{C}\quad .
$$
\item $N$-fermion states: Finally, $\left|1_N1_{N-1}\ldots1_1\right\rangle=
    \hat{\psi}_N^\dagger\hat{\psi}_{N-1}^\dagger\ldots\hat{\psi}_1^\dagger\left|0\right\rangle$ is the unique state where all the $N$ fermionic degrees of freedom are occupied because $\hat{F}\left|1_N1_{N-1}\ldots1_1\right\rangle=N\left|1_N1_{N-1}\ldots1_1\right\rangle$.

Due to the orthogonality between states with different number of fermionic degrees of freedom occupied
$$
\left\langle\left.1_{k_m}1_{k_{m-1}}\cdots1_{k_1}\right|1_{j_n}1_{j_{n-1}}\cdots1_{j_1}\right\rangle=0 \, \, \, \text {if} \, \, \, m<n \quad , \quad k_1,k_2,\ldots,k_m=1,2,\ldots,N \quad .
$$
\end{enumerate}
The fermionic Fock space has the direct sum structure written above. The dimension of $\mathcal{F}$ is the sum of the dimensions of the sub-spaces
of $\mathcal{F}_n$, $n=0,1, \ldots , N$. Thus,
$$
\text{dim}\mathcal{F}=\sum_{n=0}^N\binom{N}{n}=2^N.
$$
The Hilbert space of states of the supersymmetric quantum mechanical system is the \lq\lq direct product\rq\rq $\mathcal{SH}=L_2(\mathbb{R}^N)\otimes\mathcal{F}_{n}$ that inherits the direct sum structure of $\mathcal{F}$:
\[
\mathcal{SH}=\bigoplus_{n=0}^N\mathcal{SH}_n=\mathcal{SH}_{0}\oplus\mathcal{SH}_{1}\oplus\cdots\oplus\mathcal{SH}_{N} \, \, .
\]
The practical way to build the direct product is to allow the coefficients $f_{j_nj_{n-1}\ldots j_1}$ become square integrable functions from ${\mathbb R}^N$ to ${\mathbb C}$. Thus, the wave functions in the SUSY system are of the form:
\begin{align*}
\left.\left\langle\Psi\right|x_1,x_2,\ldots,x_N\right\rangle\left|\Psi\right\rangle=&\hspace{0.065cm}f_0(x_1,x_2,\ldots,x_N)\left|0\right\rangle+\sum_{j=1}^N f_j(x_1,x_2,\ldots,x_N)\left|1_j\right\rangle+\nonumber\\
&\hspace{0.065cm}+\sum_{\substack{j_2j_1 \\ j_2>j_1}} f_{j_2j_1}(x_1,x_2,\ldots,x_N)\left|1_{j_2}1_{j_1}\right\rangle+\cdots\nonumber\\
&\cdots+\sum_{\substack{j_nj_{n-1}\cdots j_1 \\ j_n>j_{n-1}>\cdots>j_1}} f_{j_nj_{n-1}\cdots j_1}(x_1,x_2,\ldots,x_N)\left|1_{j_n}1_{j_{n-1}}\cdots1_{j_1}\right\rangle+\cdots\nonumber\\
&\hspace{0.065cm}\cdots+\sum_{\substack{j_{N-1}j_{N-2}\cdots j_1 \\ j_{N-1}>j_{N-2}>\cdots>j_1}} f_{j_{N-1}j_{N-2}\cdots j_1}(x_1,x_2,\ldots,x_N)\left|1_{j_{N-1}}1_{j_{N-2}}\cdots1_{j_1}\right\rangle+\nonumber\\
&\hspace{0.065cm}+f_{NN-1\cdots 1}(x_1,x_2,\ldots,x_N)\left|1_N1_{N-1}\cdots1_1\right\rangle, \label{foNdim}
\end{align*}
where $f_{j_nj_{n-1}\cdots j_1}(x_1,x_2,\ldots,x_N)$ are square integrable functions.
\\
 The Fermi number operator $\hat{F}=\sum_{j=1}^N\hat{\psi}_j^\dagger\hat{\psi}_j$ and the Fermi Klein operator $\hat{K}_F=(-1)^{\hat{F}}$  label the states in the fermionic Fock space, and subsidiarily in $\mathcal{SH}$: $\hat{F}\left|F_n\right\rangle=n\left|F_n\right\rangle$ and $\hat{K}_F\left|F_n\right\rangle=(-1)^n\left|F_n\right\rangle$. Their bosonic counterparts, i.e., the Bose number operator $\hat{B}=\sum_{j=1}^N\hat{\psi}_j\hat{\psi}_j^\dagger$ and the Bose Klein operators $\hat{K}_B=(-1)^{\hat{B}}$ do the same in a dual labeling: $\hat{B}\left|F_n\right\rangle=(N-n)\left|F_n\right\rangle$ and $\hat{K}_B\left|F_n\right\rangle=(-1)^{N-n}\left|F_n\right\rangle$.
\\
In general, the operator $\hat{A}$ acting on $\mathcal{SH}$ is classified as bosonic or even if $[\hat{A},\hat{K}_F]=0$ and $\hat{A}$ is fermionic or odd if $\{\hat{A},\hat{K}_F\}=0$. Therefore, the Klein operator $\hat{K}_F$ defines a $\mathbb{Z}_2$-grading on $\mathcal{SH}$ such that $\hat{A}$ even preserves the grade in $\mathcal{SH}$ but $\hat{A}$ odd reverses this grade. In particular, the operators defining the SUSY algebra \eqref{algqqd} satisfy:
$$
[\hat{H},\hat{K}_F]=0,\ \ \ \{\hat{Q},\hat{K}_F\}=0,\ \ \ \{\hat{Q}^{\dagger},\hat{K}_F\}=0.
$$
The first commutator tells us, on one hand, that $\hat{K}_F$ is a conserved quantity and, on the other hand, that $\hat{H}$ is even. The other two anti-commutators means that $\hat{Q}$ and $\hat{Q}^{\dagger}$ are odd operators.

Because it is an even operator the supersymmetric Hamiltonian preserves the decomposition
\[
\mathcal{SH}=\bigoplus_{n=0}^N\mathcal{SH}_n=\mathcal{SH}_{0}\oplus\mathcal{SH}_{1}\oplus\cdots\oplus\mathcal{SH}_{N}
\]
acting diagonally on each component. The super-charges $\hat{Q}$ y $\hat{Q}^{\dagger}$, however, are odd operators that move states between different fermionic sectors, in such a way that the $\mathbb{Z}_2$-graded complex is built
$$
\mathcal{SH}_{0}\stackrel[\hat{Q}]{\hat{Q}^\dagger}{\rightleftarrows}\mathcal{SH}_{1}\stackrel[\hat{Q}]{\hat{Q}^\dagger}{\rightleftarrows}
\mathcal{SH}_{2}\cdots\mathcal{SH}_{N-2}\stackrel[\hat{Q}]{\hat{Q}^\dagger}{\rightleftarrows}\mathcal{SH}_{N-1}\stackrel[\hat{Q}]
{\hat{Q}^\dagger}{\rightleftarrows}\mathcal{SH}_{N}
$$
and its SUSY-cohomology defined by the kernels and images of $\hat{Q}$ and $\hat{Q^\dagger}$.
\subsection{Fermionic operators and the Clifford algebra of $\mathbb{R}^{2N}$}
Contrarily to the commutative Heisenberg algebra (\ref{relconmNdim}) which does not admit finite dimensional representations, the anti-commutative algebra (\ref{relanticonmNdim}) closed by $\hat{\psi}_j$ and $\hat{\psi}_j^\dagger$ can be represented as the Clifford algebra of $\mathbb{R}^{2N}$.  The algebra $C(\mathbb{R}^{2N})$ is generated by the Euclidean gamma matrices $\gamma^j,\ \gamma^{N+j},\ j=1,2,\ldots,N$ such that:
$$
\{\gamma^j,\gamma^k\}=2\delta^{jk}=\{\gamma^{N+j},\gamma^{N+k}\},\ \{\gamma^j,\gamma^{N+k}\}=0,\ \forall j,k=1,2,\ldots,N,
$$
$$
(\gamma^j)^\dagger=\gamma^j,\ (\gamma^{N+j})^\dagger=\gamma^{N+j}.
$$
 These $2^N\times2^N$ hermitian matrices form the irreducible representation of $C(\mathbb{R}^{2N})$ acting on the space of Euclidean spinors of dimension $2^N$.
\\
The fermionic operators are thus represented as the following combinations of these gamma matrices:
$$
\hat{\psi}_j=\frac{1}{2\sqrt{m}}(\gamma^j+i\gamma^{N+j})=\hat{\psi}_1^j+i\hat{\psi}_2^j, \ \ \ \hat{\psi}_j^\dagger=\frac{1}{2\sqrt{m}}(\gamma^j-i\gamma^{N+j})=\hat{\psi}_1^j-i\hat{\psi}_2^j
$$
The $n$-fermion state $\left|1_{j_n}1_{j_{n-1}}\ldots1_{j_1}\right\rangle$ in this Clifford algebra representation becomes the Euclidean spinor
$$
\left|1_{j_n}1_{j_{n-1}}\ldots1_{j_1}\right\rangle\rightarrow(0,\overbrace{0,\ldots,0}^{N},\ldots,\overbrace{0,\ldots,-1,\ldots,0}^{\binom{N}{n}}
,\ldots,\overbrace{0,\ldots,0}^{N},0)^T.
$$
whereas the Fermi number operator is the $2^N\times2^N$ diagonal matrix:
$$
\hat{F}=\text{diag}(0,\overbrace{1,\ldots,1}^{N},\overbrace{2,\ldots,2}^{\binom{N}{2}},\ldots,\overbrace{n,\ldots,n}^{\binom{N}{n}},
\ldots,\overbrace{N-1,\ldots,N-1}^{N},N)
$$
The Hamiltonian operator is thus block diagonal where the blocks are as follow: $\hat{H}_0=\left.\hat{H}\right|_{\mathcal{H}_0},\ \hat{H}_N=\left.\hat{H}\right|_{\mathcal{H}_N}$ are scalar Hamiltonians, $\hat{H}_1=\left.\hat{H}\right|_{\mathcal{SH}_1},\ \hat{H}_{N-1}=\left.\hat{H}\right|_{\mathcal{SH}_{N-1}}$ are $N\times N$ matrix Hamiltonian operators and $\hat{H}_n=\left.\hat{H}\right|_{\mathcal{SH}_n},\ \hat{H}_{N-n}=\left.\hat{H}\right|_{\mathcal{SH}_{N-n}}$ also $\binom{N}{n}\times\binom{N}{n}$ matrix Hamiltonian operators.

\subsection{The super-symmetric Hydrogen Atom in three dimensions}
The Wipf et al formulation of the ${\cal N}=2$ supersymmetric Hydrogen Atom in 3D space materializes the structure just described with the following choice of super-potential:
\begin{equation}
W(x_1,x_2,x_3) = -\frac{m e^2}{\hbar}\cdot r=  -\frac{m e^2}{\hbar}\cdot \sqrt{x_1^2+x_2^2+x_3^2} \label{sup3dha} \quad .
\end{equation}
The supercharges are accordingly
\[
\hat{Q}=\sum_{j=1}^3 i \left(\hbar\hat{\psi}\frac{\partial}{\partial x_j}-\frac{m e^2}{\hbar}\cdot\hat{h}\right) \quad , \quad \hat{Q}^\dagger=\sum_{j=1}^3 i \left(\hbar\hat{\psi}^\dagger\frac{\partial}{\partial x_j}-\frac{m e^2}{\hbar}\cdot\hat{h}^\dagger \right) º, \, ,
\] 
where
\[
\hat{h}= \sum_{j=1}^3\, \frac{x_j}{r}\cdot \hat{\psi}_j \quad , \quad \hat{h}^\dagger = \sum_{j=1}^3\, \frac{x_j}{r}\cdot \hat{\psi}_j^\dagger \quad , \quad m\{\hat{h},\hat{h}^\dagger\}={\mathbb I}_{2^3}
\]
are hedgehox projections and ${\mathbb I}_{2^3}$ is the $8\times 8$ unit matrix. The supersymmetric Hamiltonian is easily computed to find:
\begin{equation}
\hat{H}=\frac{1}{2m}\{\hat{Q},\hat{Q}^\dagger\}=\left(-\bigtriangleup + \frac{m e^4}{2 \hbar^2}\right)\cdot {\mathbb I}_{2^3}-\frac{e^2}{r}\cdot \hat{B} \quad , \quad \hat{B}={\mathbb I}_{2^3}-m \,\hat{F}+ m \, \hat{h}^\dagger\hat{h} \, \, \, \, . \label{suham3dah}
\end{equation}
The ${\hat H}$-operator is block diagonal acting in the irreducible components of the Hilbert space decomposition, $ L^2({\mathbb R}^3)\otimes{\cal F}_{2^3}=L^2({\mathbb R}^3)\otimes{\cal F}_0 \oplus L^2({\mathbb R}^3)\otimes{\cal F}_1 \oplus L^2({\mathbb R}^3)\otimes{\cal F}_2 \oplus L^2({\mathbb R}^3)\otimes {\cal F}_3$. Two scalar Hamiltonians act on the one-dimensional sub-spaces where the Fermi operator ${\hat F}$ evaluates to
either $0$ or $3$:
\begin{equation}
\hat{H}_0=-\frac{\hbar^2}{2 m} \bigtriangleup + \frac{m e^4}{2 \hbar^2}- \frac{e^2}{r} \quad , \quad \hat{H}_3=-\frac{\hbar^2}{2 m} \bigtriangleup + \frac{m e^4}{2 \hbar^2}+\frac{e^2}{r} \label{scsush} \, \, .
\end{equation}
The other two diagonal blocks are $3\times 3$-matrix differential operators which act respectively on $L^2({\mathbb R}^3)\otimes{\cal F}_1$ and $L^2({\mathbb R}^3)\otimes{\cal F}_2$:

\begin{equation}
\hat{H}_1=\left(\begin{array}{ccc} -\frac{\hbar^2}{2 m} \bigtriangleup + \frac{m e^4}{2\hbar^2}-e^2\cdot\frac{x_1^2}{r^3} & -e^2 \cdot\frac{x_1x_2}{r^3} & -e^2\cdot\frac{x_1x_3}{r^3} \\ -e^2\cdot\frac{x_1x_2}{r^3} & -\frac{\hbar^2}{2 m} \bigtriangleup + \frac{m e^4}{2\hbar^2}-e^2\cdot\frac{x_2^2}{r^3} & -e^2\cdot \frac{x_2x_3}{r^3} \\ -e^2\cdot\frac{x_1x_3}{r^3} & -e^2\cdot\frac{x_2x_3}{r^3} & -\frac{\hbar^2}{2 m} \bigtriangleup + \frac{m e^4}{2 \hbar^2}-e^2\cdot\frac{x_3^2}{r^3}\end{array}\right) \label{math1} \, \, \, ,
\end{equation}
\begin{equation}
\hat{H}_2=\left(\begin{array}{ccc} -\frac{\hbar^2}{2 m} \bigtriangleup + \frac{m e^4}{2\hbar^2}+e^2\cdot\frac{x_3^2}{r^3} & -e^2\cdot \frac{x_2x_3}{r^3} & e^2\cdot\frac{x_1x_3}{r^3} \\ -e^2\cdot\frac{x_2x_3}{r^3} & -\frac{\hbar^2}{2 m} \bigtriangleup + \frac{m e^4}{2\hbar^2}+e^2\cdot\frac{x_2^2}{r^3} & -e^2 \cdot\frac{x_1x_2}{r^3} \\ e^2\cdot\frac{x_1x_3}{r^3} & -e^2\cdot\frac{x_1x_2}{r^3} & -\frac{\hbar^2}{2 m} \bigtriangleup + \frac{m e^4}{2 \hbar^2}+e^2\cdot\frac{x_1^2}{r^3}\end{array}\right) \label{mathh2} \quad .
\end{equation}
Understanding the physics within this system is helped by noticing that the operators
\begin{equation}
\hat{S}_{kl}=-i\hbar m \left(\hat{\psi}_k^\dagger \hat{\psi}_l-\hat{\psi}_l^\dagger \hat{\psi}_k\right) \, \, \, , \, \, \, k,l=1,2,3
\end{equation}
represent the $SO(3)$ rotation group on the fermionic Fock space by means of the commutation rules
\begin{equation}
[\hat{S}_{kl}, \hat{S}_{ij}]= i\hbar\left(\delta_{ki}\hat{S}_{lj}+\delta_{lj}\hat{S}_{ki}-\delta_{kj}\hat{S}_{li}-\delta_{li}\hat{S}_{kj}\right) \quad , \quad [\hat{S}_{kl},\hat{\psi}_j]=i\hbar\left(\delta_{kj}\hat{\psi}_l-\delta_{lj}\hat{\psi}_k\right) \label{comrui} \quad .
\end{equation}
One can easily check that ${\hat S}_{kl}$ acts trivially in the scalar sectors ${\cal F}_0$ and ${\cal F}_3$ but ${\cal F}_1$ and ${\cal F}_2$
are adjoint representation spaces of $SO(3)$. Therefore, these three-dimensional spaces where either one or two fermionic states are occupied can be
interpreted as triplets of spin equal to one related to the $SO(3)$ spin rotations.

In fact, the total \lq\lq angular\rq\rq momentum ${\hat J}_{jl}=\hat{x}_k\hat{p}_l-\hat{x}_l\hat{p}_k +\hat{S}_{kl}$ commutes with the supersymmetric Hamiltonian,
as well as with the supercharges and Wipf et al. use this symmetry in their analysis of the spectrum. Moreover, from the structure of the block diagonal Hamiltonians
it is seen that $\hat{H}_0$ is the Schr$\ddot{\rm o}$dinger operator for one electron in the Coulomb field of a proton shifted by a constant in such a way that the ground state energy
becomes zero, henceforth the spectrum is non-negative as demanded by supersymmetry. The $\hat{H}_1$ operator, however, acts on the $L^2(\mathbb{R}^3)
\otimes{\cal F}_1$ spin one subspace. Quadrupole interactions arise between states with both
identical and different spin states, i.e. with either identical or different Fermi states occupied. Regarding the dual Hamiltonians, $\hat{H}_3$ is the Schr$\ddot{\rm o}$dinger operator governing the quantum dynamics of a positron in the field of a proton. We might say that when all the fermionic states are occupied there is transmutation from particle to antiparticle. In ${\hat H}_2$ the charge of the particle is also changed with respect to the effective charge in ${\hat H}_1$. There are also exchanges between $x_1$ and $x_3$ accompanied by sign flips except in the $x_1x_3$ combinations.

\section{${\cal N}=2$ supersymmetric quantum dynamics built from the Lieb-Liniger-Yang system of $N$ \lq\lq bosons\rq\rq on a line}

Our main task is the implementation of this supersymmetric scenario over the Lieb-Liniger-Yang exactly solvable model of $N$ \lq\lq bosons\rq\rq on a line. To buil a ${\cal N}=2$ supersymmetric quantum mechanical system having in the scalar sectors the LLY Hamiltonian,
respectively repulsive and attractive, we choose the following superpotential:
\begin{equation}
W(x_1,x_2, \ldots, x_N)=\frac{c}{2}\, \sum_{k=1}^{N-1}\, \sum_{j=k+1}^N\,|x_k-x_j|  \quad , \quad c>0 \ . \label{LLYsup}
\end{equation}
Recall that in Section \S.2 we set $\hbar=1$ and the particle masses are $m=\frac{1}{2}$. Thus, energy carries physical dimension of $L^{-2}$ and the coupling constant $c$ dimension is inverse length: $[c]=L^{-1}$. Therefore the superpotential is dimensionless and
the SUSY interactions come from the partial derivatives of $W$:
\begin{eqnarray}
&&\hspace{-0.8cm}\frac{\partial W}{\partial x_j}(x_1, \cdots, x_N)= -\frac{c}{2}\sum_{k=1}^{j-1}\varepsilon(x_k-x_j)+\frac{c}{2}\sum_{k=j+1}^N\varepsilon(x_j-x_k) \label{llsusyi1} \\ && \hspace{-0.8cm}\frac{\partial^2 W}{\partial x_j^2}(x_1, \cdots,x_N)= c \sum_{k=1}^{j-1}\delta(x_k-x_j)+c\sum_{k=j+1}^N\delta(x_j-x_k) \quad , \quad \frac{\partial^2 W}{\partial x_j\partial x_l}=-c\delta(x_j-x_l) \, \, {\rm if}\, \, j\neq l \label{llsusyi2}
\, \, .
\end{eqnarray}
where $\epsilon(z)$ is the sign function of $z$.

The configuration space of $N$ particles moving on a line is ${\mathbb R}^N$. We sall thus apply the generic formalism just described to build a supersymmetric system whose fermionic operators are extracted from the Clifford algebra of ${\mathbb R}^{2N}$ such that the choice of the superpotential (\ref{LLYsup}) and its partial derivatives given in (\ref{llsusyi1}-\ref{llsusyi2}) produce the LLY Hamiltonian
as the restriction of the supersymmetric Hamiltonian to the scalar sectors. Even though the contact interactions introduced this way break
the symmetry under the $SO(N)$ rotation group we still may characterize the Fermionic sectors ${\cal F}_1, {\cal F}_2, \cdots, {\cal F}_{N-1}$ as $SO(N)$ irreducible representations under the action of the spin operators $\hat{S}_j$. These states with one, two, three, etcetera, occupied Fermionic states can be interpreted as describing internal, or spin, degrees of freedom of the supersymmetric particles, with an internal space of states of dimension $\binom{N}{n}$. A very important CAVEAT: we are refereeing to terminology on Fermi and Bose states in two completely different frameworks. The LLY models in quantum mechanics may be applied to either boson or fermions inthe sanse
of the behaviour of the wave functions under permutations of indistinguishable $N$-particles. In supersymmetric Quantum Mechanics we have used the concept of Bose and Fermi states in the sense of Quantum Field Theory, albeit with a finite dimensional Fock space. Another difference is the following: the LLY system may be easily applied to Fermions by adapting the Bethe ansatz to Fermi statistics and adjusting
the matching conditions. There are even generalizations in the Literature enlarging the space of states to particles supporting spin. The contact interactions, however, do not affect spin. In our supersymmetric system the contact interactions arising in non-diagonal elements of the matrix differential operators leave room to spin flip after the scattering of two particles, see (\ref{llsusyi2}).

The ensuing formalism can be compared with the super-symmetric generalization of integrable systems of the Calogero-Sutherland type, see \cite{Brink}. However, our approach to this problem is intended to obtain analytical results on the supersymmetric spectrum rather than to describe in detail the algebraic structures. In fact, our choice of superpotential give rise to the LLY Hamiltonian, respectively repulsive and attractive, in the scalar sectors. Thus all the eigenfunctions of the non SUSY LLY Hamiltonian enter the spectrum of the supersymmetric Hamiltonian. The action of supercharges on eigenfunctions in the scalar sectors ${\cal S}{\cal H}_0$ and ${\cal S}{\cal H}_N$ provides eigenfunctions in the sectors $L^2(\mathbb{R}^2)\otimes {\cal F}_1$ and $L^2(\mathbb{R}^2)\otimes {\cal F}_{N-1}$. Zero modes, however, must be separately described from the scratch.

\subsection{${\cal N}=2$ supersymmetric quantum mechanics built from $N=2$ LLY bosons: Lieb-Liniger-Yang generalized SUSY interactions}

In the case of two particles, $N = 2$, the super-potential \eqref{LLYsup} simply corresponds to:
\begin{equation}
W(x_1,x_2)=\frac{c}{2}\hspace{0.05cm}|x_1-x_2| \quad ,  \quad c>0 \label{wys2p}
\end{equation}
Its derivatives include the following distributions:
\begin{equation*}
\frac{\partial W}{\partial x_1}=\frac{c}{2}\hspace{0.05cm}\varepsilon(x_1-x_2) \quad , \quad \frac{\partial W}{\partial x_2}=-\frac{c}{2}\hspace{0.05cm}\varepsilon(x_1-x_2)
\end{equation*}
\begin{equation*}
\frac{\partial^2 W}{\partial x_1^2}=\frac{\partial^2 W}{\partial x_2^2}=c\hspace{0.05cm}\delta(x_1-x_2) \quad , \quad \frac{\partial^2 W}{\partial x_1\partial x_2}=\frac{\partial^2 W}{\partial x_2\partial x_1}=-c\hspace{0.05cm}\delta(x_1-x_2) \ ,
\end{equation*}
where $\varepsilon(x)$ is the sign function and $\delta(x)$ denotes the Dirac delta function. The key elements of the supersymmetric structure, namely the supercharges, are the $4\times 4$ matrix differential operators:  \begin{align*}
\hat{Q}&=i\sum_{j=1}^2\hat{\psi}_j\hat{D}_j=i\left\{\hat{\psi}_1\left[\frac{\partial}{\partial x_1}+\frac{c}{2}\varepsilon(x_1-x_2)\right]+\hat{\psi}_2\left[\frac{\partial}{\partial x_2}-\frac{c}{2}\varepsilon(x_1-x_2)\right]\right\}\nonumber\\ &=\left(\begin{array}{cccc} 0 & \frac{\partial}{\partial x_1}+\frac{c}{2}\varepsilon(x_1-x_2) & \frac{\partial}{\partial x_2}-\frac{c}{2}\varepsilon(x_1-x_2) & 0 \\ 0 & 0 & 0 & -\frac{\partial}{\partial x_2}+\frac{c}{2}\varepsilon(x_1-x_2) \\ 0 & 0 & 0 & +\frac{\partial}{\partial x_1}+\frac{c}{2}\varepsilon(x_1-x_2) \\ 0 & 0 & 0 & 0 \end{array}\right)
\end{align*}
\begin{align*}
\hat{Q}^\dagger&=i\sum_{j=1}^2\hat{\psi}_j^\dagger\hat{D}_j^\dagger=i\left\{\hat{\psi}_1^\dagger\left[\frac{\partial}{\partial x_1}-\frac{c}{2}\varepsilon(x_1-x_2)\right]+\hat{\psi}_2^\dagger\left[\frac{\partial}{\partial x_2}+\frac{c}{2}\varepsilon(x_1-x_2)\right]\right\}\nonumber\\
&= \left(\begin{array}{cccc} 0 & 0 & 0 & 0 \\  \frac{\partial}{\partial x_1}-\frac{c}{2}\varepsilon(x_1-x_2) & 0 & 0 & 0\\ \frac{\partial}{\partial x_2}+\frac{c}{2}\varepsilon(x_1-x_2) & 0 & 0 & 0 \\ 0 &  -\frac{\partial}{\partial x_2}-\frac{c}{2}\varepsilon(x_1-x_2) &  \frac{\partial}{\partial x_1}-\frac{c}{2}\varepsilon(x_1-x_2) & 0 \end{array}\right)
\end{align*}
Thus, the scalar Hamiltonians become
\begin{equation*}
\hat{H}_0\equiv\left.\hat{H}\right|_{\mathcal{SH}_0}=-\hat{D}_1\hat{D}_1^\dagger-\hat{D}_2\hat{D}_2^\dagger=
-\frac{\partial^2 }{\partial x_1^2}-\frac{\partial^2 }{\partial x_2^2}+2c\delta(x_1-x_2)+\frac{c^2}{2} \ , \label{ham2sr}
\end{equation*}
\begin{equation*}
\hat{H}_2\equiv\left.\hat{H}\right|_{\mathcal{SH}_2}=-\hat{D}_1^\dagger\hat{D}_1-\hat{D}_2^\dagger\hat{D}_2=
-\frac{\partial^2 }{\partial x_1^2}-\frac{\partial^2 }{\partial x_2^2}-2c\delta(x_1-x_2)+\frac{c^2}{2} \ , \label{ham2sa}
\end{equation*}
while the matrix Hamiltonian is
\begin{equation*}
\hat{H}_1\equiv\left.\hat{H}\right|_{\mathcal{SH}_1}=\left( \begin{array}
[c]{cc}%
-\frac{\partial^2 }{\partial x_1^2}-\frac{\partial^2 }{\partial x_2^2}+\frac{c^2}{2} \hspace{0.1cm} & \hspace{0.1cm} 2c\delta(x_1-x_2)\\
2c\delta(x_1-x_2) \hspace{0.1cm} & \hspace{0.1cm} -\frac{\partial^2 }{\partial x_1^2}-\frac{\partial^2 }{\partial x_2^2}+\frac{c^2}{2}
\end{array} \right). \label{hammatr2dim}
\end{equation*}

\subsubsection{Collision states in the ${\cal S}{\cal H}_0$ and ${\cal S}{\cal H}_2$ sectors}

The scalar Hamiltonians $\hat{H}_0$ and $\hat{H}_2$ acting in the bosonic sectors, are equivalent to the non-supersymmetric Lieb-Liniger-Yang Hamiltonian, repulsive in the first case, attractive in the second case, and with its energy shifted by $\frac{c^2}{2}$ in both cases. The corresponding eigenfunctions associated to collision states where the two particles exchange positions are those given in \eqref{mon1}, \eqref{mon2} for the two regions $x_1<x_2$ and $x_2<x_1$. The solutions of this type exist in the two bosonic sectors ${\cal S}{\cal H}_0$ and ${\cal S}{\cal H}_2$ and their energy is simply shifted by $\frac{c^2}{2}$, having all of them non-negative energy:

\[
E^{(0)}=E^{(2)}=E+\frac{c^2}{2}=k_1^2+k_2^2+\frac{c^2}{2} \ .
\]
It is convenient to recall from Section \S.2 that for these solutions to the supersymmetric system the following property holds:
\[
S(k_2,k_1)=\frac{i(k_2-k_1)-c}{i(k_2-k_1)+c}=e^{i\theta(k_2-k_1)} \, .
\]

\subsubsection{Bound states in the ${\cal S}{\cal H}_2$: bosonic zero mode}

For these same reasons, there are bound states in sector ${\cal S}{\cal H}_2$. In fact, the dimer-type states -where no exchange occurs between the two particles, are bound states of $\hat{H}_2$ with energy
\[
E^{(2)}=P^2-\frac{c^2}{4}+P^2-\frac{c^2}{4}+\frac{c^2}{2}=2 P^2 \ .
\]
Thus, when the dimer is at rest, $P=0$, the corresponding state is a bosonic ground state of the supersymmetric system, i.e.:
\begin{itemize}
\item One first checks that
\begin{equation*}
\left(\begin{array}{cccc} 0 & \frac{\partial}{\partial x_1}+\frac{c}{2}\varepsilon(x_1-x_2) & \frac{\partial}{\partial x_2}-\frac{c}{2}\varepsilon(x_1-x_2) & 0 \\ 0 & 0 & 0 & -\frac{\partial}{\partial x_2}+\frac{c}{2}\varepsilon(x_1-x_2) \\ 0 & 0 & 0 & +\frac{\partial}{\partial x_1}+\frac{c}{2}\varepsilon(x_1-x_2) \\ 0 & 0 & 0 & 0 \end{array}\right)\left(\begin{array}{c} 0 \\ 0 \\ 0 \\  e^{-\frac{c}{2}\vert x_1-x_2\vert}\end{array}\right)=\left(\begin{array}{c} 0 \\ 0 \\ 0 \\ 0\end{array}\right)
\end{equation*}
whenever
\begin{equation*}
\left(-\frac{\partial}{\partial x_2}+\frac{c}{2}\right)e^{-\frac{c}{2}(x_1-x_2)}=0=\left(-\frac{\partial}{\partial x_2}-\frac{c}{2}\right)e^{-\frac{c}{2}(x_2-x_1)}
\end{equation*}
\begin{equation*}
\left(\frac{\partial}{\partial x_1}+\frac{c}{2}\right)e^{-\frac{c}{2}(x_1-x_2)}=0=\left(\frac{\partial}{\partial x_1}-\frac{c}{2}\right)e^{-\frac{c}{2}(x_2-x_1)} \quad .
\end{equation*}
\item It is trivially observed that also: $\hat{Q}^\dagger\Psi^{(2)}_0(x_1,x_2)= 0$.  Thus the dimer at rest is a bosonic zero mode of the system living in the ${\cal S}{\cal H}_2=L^2(\mathbb{R}^2)\otimes{\cal F}_2$ where the two fermionic degrees of freedom are occupied.
\end{itemize}
\subsubsection{The spectrum in the ${\cal S}{\cal H}_1$ fermionic sector: fermionic zero mode}

The spectral problem in the sector ${\cal S}{\cal H}_1$
\begin{equation*}
\hat{H}\Psi^{(1)}=E^{(1)}\Psi^{(1)} \quad , \quad \Psi^{(1)}(x_1,x_2)=\left( \begin{array}{c}0\\\psi_1^{(1)}(x_1,x_2) \\
\psi_2^{(1)}(x_1,x_2)\\0
\end{array} \right), \label{pehm}
\end{equation*}
reduces down to solving the spectrum of the matrix differential operator Hamiltonian $\hat{H}_1$:
\begin{equation}
\left( \begin{array}
[c]{cc}%
-\frac{\partial^2 }{\partial x_1^2}-\frac{\partial^2 }{\partial x_2^2}+\frac{c^2}{2} \hspace{0.1cm} & \hspace{0.1cm} 2c\delta(x_1-x_2)\\
2c\delta(x_1-x_2) \hspace{0.1cm} & \hspace{0.1cm} -\frac{\partial^2 }{\partial x_1^2}-\frac{\partial^2 }{\partial x_2^2}+\frac{c^2}{2}
\end{array} \right)\left( \begin{array}{c}\psi_1^{(1)}(x_1,x_2) \\
\psi_2^{(1)}(x_1,x_2)
\end{array} \right)=E^{(1)}\left( \begin{array}{c}\psi_1^{(1)}(x_1,x_2) \\
\psi_2^{(1)}(x_1,x_2)
\end{array} \right). \label{pehmexp}
\end{equation}
We observe from the structure of the Hamiltonian (\ref{pehmexp}) that contact interactions only happens when particle 1 enter the collision 
in a Fermi state and particle 2 in a Bose state exchanging the situation afterwords or viceversa. Id est, there is an spin flip interaction 
besides the $\delta$-point interaction.

This supersymmetric system is one of the special cases where there is a fermionic ground state apart from the bosonic ground state, both states being singlets of the supersymmetry algebra. In fact, spinors of the form
\[
\left( \begin{array}{c}\psi_1^{(1)}(x_1,x_2) \\
\psi_2^{(1)}(x_1,x_2)
\end{array} \right) =\left(\begin{array}{c} \psi_-(x_1,x_2) \\ -\psi_-(x_1,x_2) \end{array}\right)
\]
are eigen-spinors of the matrices $\hat{F}$ and $\hat{\Sigma}_1$:
\[
\hat{F}\Psi_-^{(1)}(x_1,x_2)=\left(\begin{array}{cccc} 0 & 0 & 0 & 0 \\ 0 & 1 & 0 & 0 \\ 0 & 0 & 1 & 0 \\ 0 & 0 & 0 & 2 \end{array}\right)\left(\begin{array}{c} 0 \\ \psi_-(x_1,x_2) \\ -\psi_-(x_1,x_2) \\ 0 \end{array}\right)=\left(\begin{array}{c} 0 \\ \psi_-(x_1,x_2) \\ -\psi_-(x_1,x_2) \\ 0 \end{array}\right)
\]
\[
 \hat{\Sigma}_1\Psi_-^{(1)}(x_1,x_2)=\left(\begin{array}{cccc} 1 & 0 & 0 & 0 \\ 0 & 0 & 1 & 0 \\ 0 & 1 & 0 & 0 \\ 0 & 0 & 0 & 1\end{array}\right)\left(\begin{array}{c} 0 \\ \psi_-(x_1,x_2) \\ -\psi_-(x_1,x_2) \\ 0 \end{array}\right) =-\left(\begin{array}{c} 0 \\ \psi_-(x_1,x_2) \\ -\psi_-(x_1,x_2) \\ 0 \end{array}\right) \, \, ,
\]
and also eigen-spinors of $\hat{H}$ with eigenvalue equal to $0$ if the following equation is satisfied
\begin{equation}
\left[-\bigtriangleup +\frac{c^2}{2}-2c\delta(x_1-x_2)\right]\psi_-(x_1,x_2)=0 \quad . \label{zmfe}
\end{equation}
Previously, it was found that the wave function $\psi_0(x_1,x_2)=-2 c e^{-\frac{c}{2}\vert x_1-x_2\vert}$, that describes a bound state of the two particles with their mass center at rest, is a solution to the equation (\ref{zmfe}). The spinor
\[
\Psi_0^{(1)}(x_1,x_2)=e^{-\frac{c}{2}\vert x_1-x_2\vert}\left(\begin{array}{c} \, \, \, \, 0 \\  \, \, \, \, 1 \\ -1 \\  \, \, \, \, 0 \end{array}\right) \quad , \quad E^{(1)}=0
\]
is therefore a fermionic ground state, i.e., it is an eigen-spinor of $\hat{H}$ with zero energy, of $\hat{\Sigma}_1$
with eigenvalue equal to $-1$, and of $\hat{F}$ with eigenvalue $1$.

We conclude that the ground states $\Psi_0^{(1)}$ and $\Psi_0^{(2)}$ form the set of zero modes of the ${\cal N}=2$ supersymmetric $N=2$ LLY system. Thus, supersymmetry is not spontaneously  broken. The Witten index
\[
n_b-n_f=\lim_{\beta\to +\infty}{\rm Tr}_{{\cal S}{\cal H}}\left[(-1)^{\hat{F}}{\rm exp}(-\beta \hat{H})\right] \ \ \ ,
\]
where $n_b$ and $n_f$ count respectively the number of bosonic and fermionic zero modes, may be not zero, even if $n_b=1$ and $n_f=1$ as in this system, because the continuous spectrum contribution comes from different spectral densities in each sector, see Reference \cite{Mosquera}.

Regarding positive energy wave functions in the sector with $\hat{F}=1$, a contact interaction problem of the Lieb-Liniger-Yang type between two particles, each of them displaying two possible states at every point in the line, must be solved. Fortunately, the structure of the supersymmetry algebra allows to obtain the eigenfunctions of the supersymmetric Hamiltonian in the sector ${\cal S}{\cal H}_1$ from the eigenfunctions of strictly positive energy in the bosonic sectors described previously. Acting on the positive energy eigenfunctions either in ${\cal S}{\cal H}_0$ with the supercharge operator $\hat{Q}^\dagger$ or in ${\cal S}{\cal H}_2$ with $\hat{Q}$ one obtains all the positive eigen-spinors of $\hat{H}_1$ living in ${\cal S}{\cal H}_1$.

\subsubsection{Fermionic collision states}

Given the wave functions,  \eqref{mon1} and \eqref{mon2}, that solve the non-supersymmetric LLY problem on the regions $R_{12}: \ x_1<x_2$ and $R_{21}: \ x_2<x_1$, respectively, whereas the $\delta$-matching  conditions during the two-particle exchange are satisfied, the corresponding fermionic states are:
\begin{equation*}
\hat{Q}^\dagger\left(\begin{array}{c} \psi_{12}(x_1,x_2) \\ 0 \\ 0 \\ 0 \end{array}\right)=\left(\begin{array}{c} 0  \\ \psi^{(1)}_{1(12)}(x_1,x_2)  \\ \psi_{2(12)}^{(1)}(x_1,x_2) \\ 0\end{array}\right) \quad , \quad \hat{Q}^\dagger\left(\begin{array}{c}  \psi_{21}(x_1,x_2) \\ 0 \\ 0 \\ 0 \end{array}\right)=\left(\begin{array}{c} 0  \\ \psi^{(1)}_{1(21)}(x_1,x_2)  \\ \psi_{2(21)}^{(1)}(x_1,x_2) \\ 0\end{array}\right) ,
\end{equation*}
\begin{equation*}
\hspace{0.075cm} \hat{Q} \hspace{0.025cm} \left(\begin{array}{c} 0 \\ 0 \\ 0 \\ \psi_{12}(x_1,x_2) \end{array}\right)=\left(\begin{array}{c} 0  \\ \psi^{(1)}_{1(12)}(x_1,x_2)  \\ \psi_{2(12)}^{(1)}(x_1,x_2) \\ 0\end{array}\right) \quad \hspace{0.1cm} , \quad \hspace{0.1cm} \hat{Q}\left(\begin{array}{c} 0 \\ 0 \\ 0 \\ \psi_{21}(x_1,x_2)  \end{array}\right)=\left(\begin{array}{c} 0  \\ \psi^{(1)}_{1(21)}(x_1,x_2)  \\ \psi_{2(21)}^{(1)}(x_1,x_2) \\ 0\end{array}\right).
\end{equation*}
A simple calculation provides the following eigenstates obtained from the monomers in the sector ${\cal S}{\cal H}_2$
through the action of $\hat{Q}$:
\begin{equation*}
\psi^{(1)}_{1(12)}(x_1,x_2)=\,\ \,\ \frac{1}{\sqrt{m}}e^{\frac{i}{2}\theta(k_1-k_2)}\left[\left( k_2-\frac{i}{2}c\right)e^{i(k_1x_1+k_2x_2)}+\left( k_1-\frac{i}{2}c\right)e^{i(k_2x_1+k_1x_2)}e^{i\theta(k_2-k_1)}\right]
\end{equation*}
\begin{equation*}
\psi^{(1)}_{2(12)}(x_1,x_2)=-\frac{1}{\sqrt{m}}e^{\frac{i}{2}\theta(k_1-k_2)}\left[\left( k_1+\frac{i}{2}c\right)e^{i(k_1x_1+k_2x_2)}+\left( k_2+\frac{i}{2}c\right)e^{i(k_2x_1+k_1x_2)}e^{i\theta(k_2-k_1)}\right]
\end{equation*}
in the region $R_{12}$ and
\begin{equation*}
\psi^{(1)}_{1(21)}(x_1,x_2)=\,\ \,\ \frac{1}{\sqrt{m}}e^{\frac{i}{2}\theta(k_1-k_2)}\left[\left( k_1+\frac{i}{2}c\right)e^{i(k_1x_1+k_2x_2)}+\left( k_2+\frac{i}{2}c\right)e^{i(k_2x_1+k_1x_2)}e^{i\theta(k_2-k_1)}\right]
\end{equation*}
\begin{equation*}
\psi^{(1)}_{2(21)}(x_1,x_2)=-\frac{1}{\sqrt{m}}e^{\frac{i}{2}\theta(k_1-k_2)}\left[\left( k_2-\frac{i}{2}c\right)e^{i(k_1x_1+k_2x_2)}+\left( k_1-\frac{i}{2}c\right)e^{i(k_2x_1+k_1x_2)}e^{i\theta(k_2-k_1)}\right]
\end{equation*}
in the region $R_{21}$. It is observed that the internal states $\psi^{(1)}_1$ and $\psi_2^{(1)}$ are exchanged when the two fermi particles interact passing from $R_{12}$ to $R_{21}$ or viceversa. Thus, the effect of the $S$-matrix over the internal degree of freedom is materialized by applying the Pauli matrix
$$
\sigma_1=\left(\begin{array}{cc} 0 & 1 \\ 1 & 0 \end{array}\right) \, \, .
$$
In this way a supersymmetric form of the Bethe ansatz arises.

The states obtained acting on the monomers in ${\cal S}{\cal H}_0$ with $\hat{Q}^\dagger$ can be calculated similarly and read:
\begin{equation*}
\psi^{(1)}_{1(12)}(x_1,x_2)=-\frac{1}{\sqrt{m}}e^{\frac{i}{2}\theta(k_1-k_2)}\left[\left( k_1-\frac{i}{2}c\right)e^{i(k_1x_1+k_2x_2)}+\left( k_2-\frac{i}{2}c\right)e^{i(k_2x_1+k_1x_2)}e^{i\theta(k_2-k_1)}\right]
\end{equation*}
\begin{equation*}
\psi^{(1)}_{2(12)}(x_1,x_2)=-\frac{1}{\sqrt{m}}e^{\frac{i}{2}\theta(k_1-k_2)}\left[\left( k_2+\frac{i}{2}c\right)e^{i(k_1x_1+k_2x_2)}+\left( k_1+\frac{i}{2}c\right)e^{i(k_2x_1+k_1x_2)}e^{i\theta(k_2-k_1)}\right]
\end{equation*}
in the region $R_{12}$ and
\begin{equation*}
\psi^{(1)}_{1(21)}(x_1,x_2)=-\frac{1}{\sqrt{m}}e^{\frac{i}{2}\theta(k_1-k_2)}\left[\left( k_2+\frac{i}{2}c\right)e^{i(k_2x_1+k_1x_2)}+\left( k_1+\frac{i}{2}c\right)e^{i(k_1x_1+k_2x_2)}e^{i\theta(k_2-k_1)}\right]
\end{equation*}
\begin{equation*}
\psi^{(1)}_{2(21)}(x_1,x_2)=-\frac{1}{\sqrt{m}}e^{\frac{i}{2}\theta(k_1-k_2)}\left[\left( k_1-\frac{i}{2}c\right)e^{i(k_2x_1+k_1x_2)}+\left( k_2-\frac{i}{2}c\right)e^{i(k_1x_1+k_2x_2)}e^{i\theta(k_2-k_1)}\right]
\end{equation*}
in the region $R_{21}$. It is of note that the eigenspinors coming from either ${\cal S}{\cal H}_0$ or ${\cal S}{\cal H}_2$ differ in the exchange of $k_1$ by $k_2$ and a relative sign.


\subsection{${\cal N}=2$ supersymmetric quantum dynamics built from $N=3$ LLY bosons: more Lieb-Liniger-Yang generalized SUSY interactions}

In the three-particle, $N=3$, LLY system the superpotential \eqref{LLYsup} reduces to:
\begin{equation}
W(x_1,x_2,x_3)=\frac{c}{2}\hspace{0.05cm}|x_1-x_2|+\frac{c}{2}\hspace{0.05cm}|x_1-x_3|+\frac{c}{2}\hspace{0.05cm}|x_2-x_3| \quad , \quad c>0 \label{wys3p}
\end{equation}
and therefore, its derivatives incorporate the following distributions:
\begin{equation*}
\frac{\partial W}{\partial x_1}=\frac{c}{2}\hspace{0.05cm} \varepsilon(x_1-x_2)+\frac{c}{2}\hspace{0.05cm}\varepsilon(x_1-x_3) \quad , \quad \frac{\partial W}{\partial x_2}=-\frac{c}{2}\hspace{0.05cm}\varepsilon(x_1-x_2)+\frac{c}{2}\hspace{0.05cm}\varepsilon(x_2-x_3)
\end{equation*}
\begin{equation*}
\frac{\partial W}{\partial x_3}=-\frac{c}{2}\hspace{0.05cm}\varepsilon(x_1-x_3)-\frac{c}{2}\hspace{0.05cm}\varepsilon(x_2-x_3) \quad , \quad \frac{\partial^2 W}{\partial x_1^2}=c\hspace{0.05cm}\delta(x_1-x_2)+c\hspace{0.05cm}\delta(x_1-x_3)
\end{equation*}
\begin{equation*}
\frac{\partial^2 W}{\partial x_2^2}=c\hspace{0.05cm}\delta(x_1-x_2)+c\hspace{0.05cm}\delta(x_2-x_3) \quad , \quad \frac{\partial^2 W}{\partial x_3^2}=c\hspace{0.05cm}\delta(x_1-x_3)+c\hspace{0.05cm}\delta(x_2-x_3)
\end{equation*}
\begin{equation*}
\frac{\partial^2 W}{\partial x_1\partial x_2}=-c\hspace{0.05cm}\delta(x_1-x_2)=\frac{\partial^2 W}{\partial x_2\partial x_1} \quad , \quad \frac{\partial^2 W}{\partial x_1\partial x_3}=-c\hspace{0.05cm}\delta(x_1-x_3)=\frac{\partial^2 W}{\partial x_3\partial x_1}
\end{equation*}
\begin{equation*}
\frac{\partial^2 W}{\partial x_2\partial x_3}=-c\hspace{0.05cm}\delta(x_2-x_3)=\frac{\partial^2 W}{\partial x_3\partial x_2} \quad .
\end{equation*}
The supercharges, are now the Clifford-type differential operators:
\begin{align}
\hat{Q}=&\hspace{0.2cm}i\sum_{j=1}^3\hat{\psi}_j\hat{D}_j=i\left\{\hat{\psi}_1\left[\frac{\partial}{\partial x_1}+\frac{c}{2}\hspace{0.05cm} \varepsilon(x_1-x_2)+\frac{c}{2}\hspace{0.05cm}\varepsilon(x_1-x_3)\right]+\right.\nonumber\\
&\left.+\hat{\psi}_2\left[\frac{\partial}{\partial x_2}-\frac{c}{2}\hspace{0.05cm}\varepsilon(x_1-x_2)+\frac{c}{2}\hspace{0.05cm}
\varepsilon(x_2-x_3)\right]+\hat{\psi}_3\left[\frac{\partial}{\partial x_3}-\frac{c}{2}\hspace{0.05cm}\varepsilon(x_1-x_3)
-\frac{c}{2}\hspace{0.05cm}\varepsilon(x_2-x_3)\right]\right\} \label{sc3DY}
\end{align}
\begin{align}
\hat{Q}^\dagger=&\hspace{0.2cm}i\sum_{j=1}^3\hat{\psi}_j^\dagger\hat{D}_j^\dagger=i\left\{\hat{\psi}_1^\dagger\left[\frac{\partial}{\partial x_1}-\frac{c}{2}\hspace{0.05cm} \varepsilon(x_1-x_2)-\frac{c}{2}\hspace{0.05cm}\varepsilon(x_1-x_3)\right]+\right.\nonumber\\
&\left.+\hat{\psi}_2^\dagger\left[\frac{\partial}{\partial x_2}+\frac{c}{2}\hspace{0.05cm}\varepsilon(x_1-x_2)-\frac{c}{2}\hspace{0.05cm}
\varepsilon(x_2-x_3)\right]+\hat{\psi}_3^\dagger\left[\frac{\partial}{\partial x_3}+\frac{c}{2}\hspace{0.05cm}\varepsilon(x_1-x_3)
+\frac{c}{2}\hspace{0.05cm}\varepsilon(x_2-x_3)\right]\right\} \label{sc3DY1} \, \, \, .
\end{align}
As a consequence, the scalar Hamiltonians acting on $\mathcal{SH}_0$ and $\mathcal{SH}_3$ become respectively the repulsive and attractive three particle LLY Hamiltonians
displaced in a positive constant in such a way that their spectra are non negative:
\begin{align*}
\hat{H}_0\equiv\left.\hat{H}\right|_{\mathcal{SH}_0}&=-\hat{D}_1\hat{D}_1^\dagger-\hat{D}_2\hat{D}_2^\dagger-\hat{D}_3\hat{D}_3^\dagger\nonumber\\
&=-\bigtriangleup+2c\hspace{0.05cm}\delta(x_1-x_2)+2c\hspace{0.05cm}\delta(x_1-x_3)+2c\hspace{0.05cm}\delta(x_2-x_3)+2c^2 \ ,
\end{align*}
\begin{align*}
\hat{H}_3\equiv\left.\hat{H}\right|_{\mathcal{SH}_3}&=-\hat{D}_1^\dagger\hat{D}_1-\hat{D}_2^\dagger\hat{D}_2-\hat{D}_3^\dagger\hat{D}_3\nonumber\\
&=-\bigtriangleup-2c\hspace{0.05cm}\delta(x_1-x_2)-2c\hspace{0.05cm}\delta(x_1-x_3)-2c\hspace{0.05cm}\delta(x_2-x_3)+2c^2 \ .
\end{align*}
There is a subtle but important point in the previous calculation that is worth mentioning: the cross products of the signum functions associated to the relative positions of each pair of particles always add to unity regardless of the order of the three particles over the axis:
\[
\varepsilon(x_1-x_2)\varepsilon(x_1-x_3)+ \varepsilon(x_2-x_1)\varepsilon(x_2-x_3)+\varepsilon(x_1-x_3)\varepsilon(x_2-x_3)=1 \quad .
\]
\\
The $3\times3$ matrix Hamiltonian $\hat{H}_1$ that acts in the three-dimensional $\hat{F}=1$ Fermionic Fock subspace $\mathcal{SH}_1$  is therefore:
\begin{equation}
\hat{H}_1\equiv\left.\hat{H}\right|_{\mathcal{SH}_1}=
\left( \begin{array}
[c]{ccc}%
-\bigtriangleup +2c\hspace{0.05cm}\delta(x_2-x_3)+ 2c^2 & 2c\hspace{0.05cm}\delta(x_1-x_2) & 2c\hspace{0.05cm}\delta(x_1-x_3)\\
2c\hspace{0.05cm}\delta(x_1-x_2) & -\bigtriangleup+2c\hspace{0.05cm}\delta(x_1-x_3)+2c^2 & 2c\hspace{0.05cm}\delta(x_2-x_3)\\
2c\hspace{0.05cm}\delta(x_1-x_3) & 2c\hspace{0.05cm}\delta(x_2-x_3)  & -\bigtriangleup+2c\hspace{0.05cm}\delta(x_1-x_2)+2 c^2\\
\end{array} \right)  \label{3matf1}.
\end{equation}
The physical meaning of the interactions encoded in the Hamiltonian (\ref{3matf1}) is awkward. The wave functions in $L^2(\mathbb{R}^2)\otimes{\cal F}_1$
\[
\Psi^{(1)}(x_1,x_2,x_3)=\left(\begin{array}{c} \psi^{(1)}_1(x_1,x_2,x_3)\\ \psi^{(1)}_2(x_1,x_2,x_3)\\ \psi^{(1)}_3(x_1,x_2,x_3) \end{array}\right)\in L^2(\mathbb{R}^3)\otimes {\cal F}_1
\]
three-component spinors: $\psi^{(1)}_1(x_1,x_2,x_3)$, $\psi_2^{(1)}(x:1,x_2,x_3)$, and $\psi^{(1)}_1(x_1,x_2,x_3)$ respectively describe
states where the fermionic state of the first, second, third particles is occuppied whereas the other two are empty.
Matrix elements of the Hamiltonian between $\psi^{(1)}_1(x_1,x_2,x_3)$-$\psi^{(1)}_1(x_1,x_2,x_3)$ components contain only $\delta$-interactions between the last two particles. The non-diagonal terms in the first row, the matrix elements $\psi^{(1)}_1(x_1,x_2,x_3)$-$\psi^{(1)}_2(x_1,x_2,x_3)$ and $\psi^{(1)}_1(x_1,x_2,x_3)$-$\psi^{(1)}_3(x_1,x_2,x_3)$, show contact interactions between particles with different occupation number of the Fermi states. Clearly, the same observations can be cyclically stated about wave functions where either particle 2 travels with its Fermi state occupied, $\psi^{(1)}_2(x_1,x_2,x_3)$, or, it is particle 3 the dressed with a Fermi state: $\psi^{(1)}_3(x_1,x_2,x_3)$.
The $3\times3$ matrix Hamiltonian $\hat{H}_2$ that acts in the three-dimensional $\hat{F}=2$ Fermionic Fock subspace $\mathcal{SH}_2$ is:
\begin{equation}
\hat{H}_2\equiv\left.\hat{H}\right|_{\mathcal{SH}_2}=
\left( \begin{array}
[c]{ccc}%
-\bigtriangleup -2c\hspace{0.05cm}\delta(x_1-x_2)+2 c^2& 2c\hspace{0.05cm}\delta(x_2-x_3) & -2c\hspace{0.05cm}\delta(x_1-x_3)\\
2c\hspace{0.05cm}\delta(x_2-x_3) & -\bigtriangleup-2c\hspace{0.05cm}\delta(x_1-x_3)+2c^2 & 2c\hspace{0.05cm}\delta(x_1-x_2)\\
-2c\hspace{0.05cm}\delta(x_1-x_3) & 2c\hspace{0.05cm}\delta(x_1-x_2)  & -\bigtriangleup-2c\hspace{0.05cm}\delta(x_2-x_3)+2 c^2\\
\end{array} \right) \label{hm23DY} \ \ .
\end{equation}
The wave functions 
\[
\Psi^{(2)}(x_1,x_2,x_3)=\left(\begin{array}{c} \psi^{(2)}_1(x_1,x_2,x_3)\\ \psi^{(2)}_2(x_1,x_2,x_3)\\ \psi^{(2)}_3(x_1,x_2,x_3) \end{array}\right)\in L^2(\mathbb{R}^3)\otimes {\cal F}_2
\]
characterize states where two of the three particle Fermi states are occupied and the third is empty in each two-combination of the three particles. With this idea
in the back of the mind the interactions comig from $\hat{H}_2$ can be understood along similar lines as those due to $\hat{H}_1$. It is worth, however, the comparison with the relative structure of the matricial Hamiltonians of the 3D supersymmetric Hydrogen atom (\ref{math1}) and (\ref{mathh2}). Like in the atomic case interactions change signs, from repulsive to attarctive in this case, in  
(\ref{hm23DY}) with respect to (\ref{3matf1}) except between particles 1 and 3.

\subsubsection{Collision states in the ${\cal S}{\cal H}_0$ and ${\cal S}{\cal H}_3$ sectors}

The scalar Hamiltonians  $\hat{H}_0$ and $\hat{H}_3$ are just the $N=3$ non-supersymmetric Lieb-Liniger-Yang Hamiltonians respectively repulsive and attractive with energy shifted by $2c^2$ in both cases. The eigenfunctions that correspond to collision states, where particles located at $x_1$ and $x_2$ exchange positions, are given for the two regions, $R_{123}$ and $R_{213}$, by:
\begin{equation}
\psi_{123}^{(\iota)}(x_1,x_2,x_3)=e^{\frac{i}{2}[\theta(k_1-k_2)+\theta(k_1-k_3)+\theta(k_2-k_3)]}\left[e^{i(k_1 x_1+k_2 x_2)}+e^{i(k_1 x_2+k_2 x_1)+i\theta(k_2-k_1)}\right]e^{ik_3x_3}  \label{ab2swf1}
\end{equation}
\begin{equation}
\psi_{213}^{(\iota)}(x_1,x_2,x_3)=e^{\frac{i}{2}[\theta(k_1-k_2)+\theta(k_1-k_3)+\theta(k_2-k_3)]}\left[e^{i(k_1 x_2+k_2 x_1)}+e^{i(k_1 x_1+k_2 x_2)+i\theta(k_2-k_1)}\right]e^{ik_3x_3}  \label{ab2swf2}
\end{equation}
where $\iota=0,3$ indicates the sector $\mathcal{SH}_0$ or $\mathcal{SH}_3$ where the Bethe ansatz eigenfunction lives.
\\
The solution on the region $R_{123}$, $x_1<x_2<x_3$, corresponds to the product of a plane wave $e^{ik_3x_3}$ -describing the independent movement of the particle at $x_3$- and a linear combination of two terms: (1) $e^{i(k_1 x_1+k_2 x_2)}$ where the particles at  $x_1$ and $x_2$ propagate as plane waves with momenta $k_1$ and $k_2$, and (2)  $e^{i(k_1 x_2+k_2 x_1)+i\theta(k_2-k_1)}$, where now the momenta have been exchanged according to Bose statistics. A relative phase then emerges as a result of the exchange.
On the region $R_{213}$, $x_2<x_1<x_3$, the solution is similar, with exchanged momenta $k_1$ and $k_2$, the matching conditions at the boundary between the two regions govern the exchange phase. In this case, the reduction of the collision problem for particles at $x_1$ and $x_2$ with respect to the mass center allows a reinterpretation of the exchange interaction as the phase shift induced by the Dirac delta potential $\delta(x_1-x_2)$ that applies when  the particle at $x_1$ surpasses the other particle, at $x_2$ across the boundary between $R_{123}$ and $R_{213}$ \cite{Ladislav}. During this process, the total momentum and energy are conserved quantities, thus there is elastic scattering such that the particles at  $x_1$ and $x_2$ only exchange their momenta. If the initial momenta on $R_{123}$ are $k_1$, $k_2$ and $k_3$, then the initial momenta on $R_{213}$ are $k_1^\prime=k_2$, $k_2^\prime=k_1$ and $k_3^\prime=k_3$ \cite{Ladislav}. The solution on other regions, $R_Q$ and $R_P$, where the other two pairs of particles are the ones leading the exchange, is similar.
\\
Solutions of this type exist on the two scalar sectors, and their energy is simply shifted by $2c^2$ with respect to the non SUSY energy:
\[
E^{(0)}=E^{(3)}=E+2c^2=k_1^2+k_2^2+k_3^2+2c^2 \ .
\]
It is worth pointing out that in these solutions to the supersymmetric system the following phase shifts arise:
\begin{eqnarray*}
S(k_1,k_2)&=&\frac{i(k_2-k_1)-c}{i(k_2-k_1)+c}=e^{i\theta(k_2-k_1)} \, \, \, \quad  , \, \, \, \quad
S(k_1,k_3)=\frac{i(k_3-k_1)-c}{i(k_3-k_1)+c}=e^{i\theta(k_3-k_1)} \\
S(k_2,k_3)&=&\frac{i(k_3-k_2)-c}{i(k_3-k_2)+c}=e^{i\theta(k_3-k_2)} \ ,
\end{eqnarray*}
implying that three-body scattering factors into two-particle processes.

\subsubsection{Bound states in ${\cal S}{\cal H}_3$: fermionic zero mode}

In the sector ${\cal S}{\cal H}_3$ the scalar Hamiltonian $\hat{H}_3$ accommodates attractive $\delta$ interactions when two of the three particles coincide leaving room to bound states. The bound eigen-states of $\hat{H}_3$, with Fermi number $\hat{F}=3$ odd, are those where two of the particles experience no exchange, but this pair collides with the third particle, namely a monomer-dimer type of state, and/or states where the three particles maintain their order, namely a trimer state type. The monomer-dimer states are obtained from the selection of momenta
\begin{equation}
k_1=P+i\frac{c}{2} \quad , \quad k_2=P-i\frac{ c}{2} \quad , \quad k_3=Q \quad , \quad P\neq Q\in \mathbb{R} \label{21s}
\end{equation}
and given by the Bethe ansatz wave function:
\begin{equation*}
\psi_{\rm PQ}(x_1,x_2,x_3)= e^{2i P X_{12}}e^{-\frac{ c}{2}\vert x_1-x_2\vert} e^{iQ x_3}+
e^{2i P X_{13}}e^{-\frac{c}{2}\vert x_1-x_3\vert}e^{iQ x_2}+
e^{2i P X_{23}}e^{-\frac{c}{2}\vert x_2-x_3\vert}e^{iQ x_1} \, \, ,
\end{equation*}
where $X_{ab}=(x_a+x_b)/2$ denotes the center of mass coordinate of the $(x_a,x_b)$ pair. These monomer-dimer bound states of $\hat{H}_3$ have energy:
\begin{equation}
E_{\rm m/d}^{(3)}=P^2-\frac{c^2}{4}+P^2-\frac{c^2}{4}+Q^2+2c^2=Q^2+2P^2+\frac{3c^2}{2} \label{edms3p} \ \ .
\end{equation}
The trimer-type bound states are obtained by choosing the complex momenta
\begin{equation}
k_1=P+ic \quad , \quad k_2=P \quad , \quad k_3=P-ic \quad , \quad P\in\mathbb{R} \label{3s}
\end{equation}
and given by the Bethe ansatz wave function
\begin{equation}
\psi_{\rm P}(x_1,x_2,x_3)=e^{3iPX}e^{-\frac{c}{2}\left(|x_1-x_2|+|x_1-x_3|+|x_2-x_3|\right)} \quad , 
\end{equation}
where $X=(x_1+x_2+x_3)/3$ is the center of mass coordinate of the three particles.
These bound states of $\hat{H}_3$ have energies:
\begin{equation}
E_{\rm t}^{(3)}=P^2-c^2+P^2+P^2-c^2+2c^2=3P^2 \label{ets3p}
\end{equation}
In \eqref{edms3p} one sees that $E_{\rm m/d}^{(3)}>0$, even for the case where the monomer and the dimer remain at rest. \eqref{ets3p}, however, reveals that a trimer at rest, $P=0$, has zero energy.
\\
Thus, the trimer at rest is the ground state of the supersymmetric system with $\hat{F}=3$. Since it has zero energy, it is a singlet of the supersymmetry algebra and there is no spontaneous supersymmetry breaking in the ${\cal N}=2$ SUSY $N=3$ LLY system. The eigen-spinor can be easily found by recalling that $\psi_0(x_1,x_2,x_3)=e^{-\frac{c}{2}(|x_1-x_2|+|x_1-x_3|+|x_2-x_3|)}$ is the ground state of the non-SUSY attractive Hamiltonian of LLY for three particles. The associated spinor in ${\cal S}{\cal H}_3$ belongs to the kernel of the supercharge $\hat{Q}$ \eqref{sc3DY}, $\hat{Q}\Psi_0^{(3)}(x_1,x_2,x_3)=0$. Skipping the details, it is easy to show that
\begin{equation}
\Psi_0^{(3)}(x_1,x_2,x_3)=e^{-\frac{c}{2}(|x_1-x_2|+|x_1-x_3|+|x_2-x_3|)}
\end{equation}
is a zero mode of the supersymmetric LLY system of three particles living in the ${\cal S}{\cal H}_3$ sector, henceforth carrying fermionic number $\hat{F}=3$. One must follow the steps given to find the bosonic zero mode of the susy LLY system of two particles.

\subsubsection{The spectrum in the $\hat{F}=1$ ${\cal S}{\cal H}_1$ and $\hat{F}=2$ ${\cal S}{\cal H}_2$ sectors}

The spectral problem on sector ${\cal S}{\cal H}_1$
\begin{equation*}
\hat{H}_1\Psi^{(1)}=E^{(1)}\Psi^{(1)} 
\end{equation*}
\normalsize
reduces down to solving the spectrum of the matrix Hamiltonian $\hat{H}_1$:
\begin{equation*}
\hat{H}_1\left(
\begin{array}{c}
 \psi_1^{(1)}(x_1,x_2,x_3)  \\
 \psi_2^{(1)}(x_1,x_2,x_3)  \\
 \psi_3^{(1)}(x_1,x_2,x_3)
\end{array}
\right)=E^{(1)}\left(
\begin{array}{c}
 \psi_1^{(1)}(x_1,x_2,x_3)  \\
 \psi_2^{(1)}(x_1,x_2,x_3)  \\
 \psi_3^{(1)}(x_1,x_2,x_3)
\end{array}
\right) \label{pehmexp1}
\end{equation*}
while the spectral problem on sector ${\cal S}{\cal H}_2$
\begin{equation*}
\hat{H}_2\Psi^{(2)}=E^{(2)}\Psi^{(2)} 
\end{equation*}
\normalsize
consists in finding the eigenvalues and eigenspinors of the matrix Hamiltonian $\hat{H}_2$:
\begin{equation*}
\hat{H}_2\left(
\begin{array}{c}
 \psi_1^{(2)}(x_1,x_2,x_3)  \\
 \psi_2^{(2)}(x_1,x_2,x_3)  \\
 \psi_3^{(2)}(x_1,x_2,x_3)
\end{array}
\right)=E^{(2)}\left(
\begin{array}{c}
 \psi_1^{(2)}(x_1,x_2,x_3)  \\
 \psi_2^{(2)}(x_1,x_2,x_3)  \\
 \psi_3^{(2)}(x_1,x_2,x_3)
\end{array}
\right). \label{pehmexp2}
\end{equation*}
Both Hamiltonians are generalizations of the LLY interactions among three particles, with \lq\lq spin\rq\rq equal to $1$, that move on a line but whose contact interactions involve not only positions but also spin states. A direct analytical solution is not possible, the Bethe ansatz does not work in this situation because there are \lq\lq spin\rq\rq flips in two-particle exhanges due to the non-diagonal matrix elements in $\hat{H}_1$ and $\hat{H}_2$. However, the structure of the supersymmetric algebra permits finding half of the eigenfunctions of the supersymmetric Hamiltonian in sectors ${\cal S}{\cal H}_1$ and ${\cal S}{\cal H}_2$ based on the eigenfunctions already known in the scalar sectors. Since Hamiltonians $\hat{H}_0$ and $\hat{H}_3$ are of standard LLY type, by applying the supercharge operators $\hat{Q}$ and $\hat{Q}^\dagger$ to their eigenfunctions one obtains eigenspinors of $\hat{H}$ respectively in $\mathcal{SH}_1$ and $\mathcal{SH}_2$.

\subsubsection{Bosonic $\hat{F}=2$ ground state in ${\cal S}{\cal H}_2$ and bound states of monomer-dimer-type}

Because analyzing the variety of ground states in any supersymmetric system is of outmost importance regarding the crucial problem of supersymmetry spontaneous breaking, we now explre the existence of a second ground state, in this case a bosonic state with fermionic number $\hat{F}=2$, which should be added to the already identified ground state with $\hat{F}=3$. To prove this statement we first observe that, on eigenspinors of the matrix $\hat{\Sigma}_1^{(2)}$ with eigenvalue $-1$,
\begin{equation*}
\hat{\Sigma}_1^{(2)} \Psi_-^{(2)}(x_1,x_2,x_3)=\frac{1}{2}\left(\begin{array}{ccc} \, \, \, 0 & 1 & -1 \\ \, \, \, 1 & 0 & \, \, \, 1 \\ -1 & 1 & \, \, \, 0\end{array}\right)\left(\begin{array}{c}\, \, \, \psi_-(x_1,x_2,x_3) \\-\psi_-(x_1,x_2,x_3) \\ \, \, \, \psi_-(x_1,x_2,x_3)\end{array}\right)=-\left(\begin{array}{c}\, \, \, \psi_-(x_1,x_2,x_3) \\-\psi_-(x_1,x_2,x_3) \\ \, \, \, \psi_-(x_1,x_2,x_3)\end{array}\right) \, \, ,
\end{equation*}
it is easily checked that
\begin{equation*}
\hspace{-0.2cm}\left( \begin{array}
[c]{ccc}%
-\bigtriangleup -2c\hspace{0.05cm}\delta(x_1-x_2)+2 c^2& 2c\hspace{0.05cm}\delta(x_2-x_3) & -2c\hspace{0.05cm}\delta(x_1-x_3)\\
2c\hspace{0.05cm}\delta(x_2-x_3) & -\bigtriangleup-2c\hspace{0.05cm}\delta(x_1-x_3)+2c^2 & 2c\hspace{0.05cm}\delta(x_1-x_2)\\
-2c\hspace{0.05cm}\delta(x_1-x_3) & 2c\hspace{0.05cm}\delta(x_1-x_2)  & -\bigtriangleup-2c\hspace{0.05cm}\delta(x_2-x_3)+2 c^2\\
\end{array} \right)\left(\begin{array}{c} \, \, \, \, \, \psi_-(\vec{x}) \\ -\psi_-(\vec{x}) \\ \, \, \, \, \, \psi_-(\vec{x})\end{array}\right) \, \, \,
\end{equation*}
is equivalent to:
\begin{equation*}
\left[ -\bigtriangleup -2c\hspace{0.05cm}\delta(x_1-x_2) -2c\hspace{0.05cm}\delta(x_1-x_3)-2c\hspace{0.05cm}\delta(x_2-x_3)+2c^2\right]\psi_-(x_1,x_2,x_3) \, \, \, .
\end{equation*}
Therefore, the fermionic zero mode $\Psi_0^{(3)}(x_1,x_2,x_3)$ described above is acompanied by other zero mode, this time bosonic,
given by the spinor
\begin{equation*}
\Psi^{(2)}_0(x_1,x_2,x_3)=e^{-\frac{|c|}{2}(|x_1-x_2|+|x_1-x_3|+|x_2-x_3|)}\left(\begin{array}{c} \, \, \, \, 1 \\ -1 \\ \, \, \, \, 1 \end{array}\right)
\end{equation*}
which is then a ground state of $\hat{H}_2$, eigen-spinor of both $\hat{F}$, with eigenvalue equal to $2$, and $\hat{\Sigma}_1^{(1)}$
with eigenvalue equal to $-1$. The structure of the variety of ground states is then identical for the supersymmetric LLY models with $N=3$ and $N=2$ particles. In both cases, there is a bosonic ground state as well as a fermionic ground state. The supersymmetry of the system does not break spontaneously. The Witten index, the number of bosonic minus the number of fermionic ground states is zero if one discards
possible contributions due to Fermi-Bose spectral asymmetries in the continuous spectrum. The only difference between the $N=2$ and $N=3$ cases is that the bosonic ground state belongs to a scalar sector and the fermionic ground state belongs to a \lq\lq spinorial\rq\rq sector if $N=2$, whereas in the three-particle system a converse situation occurs: namely the bosonic ground state lies in a \lq\lq spinorial\rq\rq sector but the fermionic ground state lies in a scalar sector.
\\
Clearly, in sector ${\cal S}{\cal H}_2$, there are bosonic bound states of $\hat{H}_2$ resulting from existing fermionic trimers and monomer-dimers in ${\cal S}{\cal H}_3$:
\begin{eqnarray*}
\Psi^{(2)}_{P}(x_1,x_2,x_3)&=&\hat{Q}\Psi^{(3)}_{P}(x_1,x_2,x_3) \\ \Psi^{(3)}_{P}(x_1,x_2,x_3)&=&\psi_{P}(x_1,x_2,x_3)\left(\begin{array}{cccccccc} 0 & 0 & 0 & 0 & 0 & 0 & 0 & 1\end{array}\right)^{\rm T}
\end{eqnarray*}
where
\begin{equation*}
\psi_{P}(x_1,x_2,x_3)=e^{i P X}e^{-\frac{c}{2}\left(\vert x_1-x_2\vert+\vert x_1-x_3\vert+\vert x_2-x_3\vert\right)}
\end{equation*}
and
\begin{eqnarray*}
\Psi^{(2)}_{QP}(x_1,x_2,x_3)&=&\hat{Q}\Psi^{(3)}_{QP}(x_1,x_2,x_3) \\ \Psi^{(3)}_{QP}(x_1,x_2,x_3)&=&\psi_{QP}(x_1,x_2,x_3)\left(\begin{array}{cccccccc} 0 & 0 & 0 & 0 & 0 & 0 & 0 & 1\end{array}\right)^{\rm T}
\end{eqnarray*}
where
\begin{equation*}
\psi_{\rm PQ}(x_1,x_2,x_3)=e^{2i P X_{12}}e^{-\frac{\vert c\vert}{2}\vert x_1-x_2\vert}e^{iQ x_3}+
e^{2i P X_{13}}e^{-\frac{\vert c\vert}{2}\vert x_1-x_3\vert}e^{iQ x_2}+
e^{2i P X_{23}}e^{-\frac{\vert c\vert}{2}\vert x_2-x_3\vert }e^{iQ x_1} \, \, .
\end{equation*}

ple, but demanding, to calculate these eigenfunctions on the other regions.

\subsection{${\cal N}=2$ supersymmetric quantum dynamics built from ${\bf N}$ LLY bosons on a line: all the generalized Lieb-Liniger-Yang SUSY interactions}
The derivatives of the superpotential (\ref{LLYsup}) up to second order
\begin{eqnarray}
&& \frac{\partial W}{\partial x_j}=-\frac{c}{2}\cdot\left(\sum_{k=1}^{j-1}\, \varepsilon(x_k-x_j)-\sum_{k=j+1}^N\, \varepsilon(x_j-x_k)\right) \quad , \quad \frac{\partial^2 W}{\partial x_j^2}=-c\cdot\sum_{k\neq j}\, \delta(x_j-x_k) \label{diag LLY}\\ &&\frac{\partial^2 W}{\partial x_j \partial x_k}=c\delta(x_j-x_k)=\frac{\partial^2 W}{\partial x_k \partial x_j} \label{ndiag LLY}
\end{eqnarray}
encode all the contact interactions compatible with supersymmetry in the generalized ${\cal N}=2$ SUSY Lieb-Liniger-Yang model.
In the scalar sectors $\mathcal{SH}_0$ and $\mathcal{SH}_N$, where the eigenvalues of the Fermi number operator $\hat{F}$ are respectively
$0$ and $N$, the supersymmetric LLY Hamiltonian reduces to the Schr$\ddot{\rm o}$dinger operators
\begin{eqnarray}
\hat{H}_0=\hat{H}\Big\vert_{\mathcal{SH}_0}&=&-\bigtriangleup+2c\sum_{k=1}^{N-1}\, \sum _{j=k+1}^N\,\delta(x_k-x_j)+\frac{c^2}{12}N(N^2-1) \label{0hamlly}\\ \hat{H}_N=\hat{H}\Big\vert_{\mathcal{SH}_N}&=&-\bigtriangleup-2c\sum_{k=1}^{N-1}\, \sum _{j=k+1}^N\,\delta(x_k-x_j)+\frac{c^2}{12}N(N^2-1)\label{Nhamlly}\\ \bigtriangleup &=& \frac{\partial^2}{\partial x_1^2}+ \frac{\partial^2}{\partial x_2^2}+\frac{\partial^2}{\partial x_3^2}+ \cdots +\frac{\partial^2}{\partial x_N^2} \quad .
\end{eqnarray}
Both (\ref{0hamlly}) and (\ref{Nhamlly}) are standard Lieb-Liniger-Yang Hamiltonians for $N$ particles displaced in a constant guaranteeing that the ground state energy is non negative. The contact interactions in $\hat{H}_0$ are repulsive in such a way that only collision states appear in the $\hat{H}_0$-spectrum. All the point interactions in $\hat{H}_N$, however, are attractive leaving room to  bound states
of different types. The ground state of $\hat{H}_N$ is the $N$-mer at rest
\[
\psi_0^{(N)}(x_1,x_2,x_3)={\rm exp}\left(-\frac{c}{2}\cdot \sum_{k=1}^{N-1}\, \sum_{j=k+1}^N \, \vert x_k-x_j\vert\right)\, \, \, ,
\]
which is a, bosonic if $N$ is even, fermionic if $N$ is odd, zero mode of $\hat{H}$. ${\cal N}=2$ supersymmetry is thus unbroken.

There is another ground state of the system in the sector characterized by a \lq\lq fermionic\rq\rq number $\hat{F}=N-1$. The operator $\hat{H}_{N-1}=\hat{H}\Big\vert_{\mathcal{SH}_{N-1}}$ is an $N\times N$-matrix differential operator because ${\rm dim}\, {\cal F}_{N-1}=\left(\begin{array}{c} N \\ N-1\end{array}\right)=N$. Acting on $N$-dimensional \lq\lq spinors\rq\rq of an special type one checks that:
\begin{equation*}
\hspace{-0.1cm}\hat{H}_{N-1} \left(\begin{array}{c}\psi_-(x_1, \ldots ,x_N)\\ -\psi_-(x_1, \ldots ,x_N) \\ \psi_-(x_1, \ldots ,x_N) \\ \vdots \\ (-1)^N \psi_-(x_1, \ldots ,x_N) \end{array}\right) \,\,   \equiv \, \,  \left(-\bigtriangleup-2c\sum_{k=1}^{N-1}\, \sum _{j=k+1}^N\,\delta(x_k-x_j)+\frac{c^2}{12}N(N^2-1)\right) \psi_-(x_1, \ldots ,x_N) \,  .
\end{equation*}
Thus,
\[
\psi_0^{(N-1)}(x_1,x_2,x_3)={\rm exp}\left(-\frac{c}{2}\cdot \sum_{k=1}^{N-1}\, \sum_{j=k+1}^N \, \vert x_k-x_j\vert\right)\cdot \left(\begin{array}{c} \, \, \, \, \, 1 \\ -1 \\ \, \, \, \, \, 1 \\ \, \, \, \, \vdots \\ \, \, \, (-1)^N \end{array}\right)
\]
is the second zero mode of the system, fermionic if $N$ is even, bosonic if $N$ is odd. Again, modulo subleties due to the continuous spectra, the Witten index measuring the difference between
bosonic and fermionic ground states, is zero despite supersymmetry remaining unbroken, which always happen if the ground state is a zero mode, after the choice of ground state. Other eigenstates of $\hat{H}$ may be analitycally identified. By means of the action of $\hat{Q}^\dagger$ and $\hat{Q}$ on the already described collision states in $\hat{H}_0$ and $\hat{H}_N$, see Section \S.2, new collision eigenstates of $\hat{H}$, respectively living in $\mathcal{SH}_1$ and $\mathcal{SH}_{N-1}$, are obtained. Of course, all bound state types existing in $\mathcal{SH}_N$, from those binding only one pair of particles to moving $N$-mers, give rise to new bound states in $\mathcal{SH}_{N-1}$ simply under the $\hat{Q}$-action on them.

Finally, we briefly comment on the promotion to a supersymmetric status of the LLY system of $N$ bosons moving on the finite
interval $[0,L]$. The main novelty with respect to the non SUSY system is to realize that cyclic or periodic boundary conditions are not compatible with supersymmetry. SUSY preserving boundary conditions are established in two steps:
\begin{enumerate}

\item First, Dirichlet boundary conditions are imposed on any generic spinor:
\begin{equation}
\Psi^{(n)}(x_1,x_2, \ldots, x_j=0, \ldots x_n)=\Psi^{(n)}(x_1,x_2, \ldots, x_j=L, \ldots x_n)=0 \, \, \, , \, \forall j=1,2, \ldots, N \, .
\label{dirbcs}
\end{equation}
$n=0,1, \ldots, N$ is the Fermi number of the $\left(\begin{array}{c} N \\ n\end{array}\right)$-component \lq\lq spinor\rq\rq: $\hat{F}\Psi^{(n)}=n\, \Psi^{(n)}$.

\item Second, if $\Psi^{(n)}\neq \hat{Q}^\dagger \Psi^{(n-1)}$, the SUSY partner $\Psi^{(n+1)}=\hat{Q}^\dagger \Psi^{(n)}$
is obliged to comply with Robin boundary conditions:
\begin{equation}
\hat{Q}^\dagger\Psi^{(n)}(x_1,x_2, \ldots, x_j=0, \ldots x_n)=\hat{Q}^\dagger\Psi^{(n)}(x_1,x_2, \ldots, x_j=L, \ldots x_n)=0 \, \, \, , \, \forall j=1,2, \ldots, N \, .
\label{robbcs}
\end{equation}
Alternatively, if $\Psi^{(n)}\neq \hat{Q}\Psi^{(n+1)}$, the SUSY partner $\Psi^{(n-1)}=\hat{Q} \Psi^{(n)}$
complies also with Robin boundary conditions:
\begin{equation}
\hat{Q}\Psi^{(n)}(x_1,x_2, \ldots, x_j=0, \ldots x_n)=\hat{Q}\Psi^{(n)}(x_1,x_2, \ldots, x_j=L, \ldots x_n)=0 \, \, \, , \, \forall j=1,2, \ldots, N \, .
\label{robbcs1}
\end{equation}
\end{enumerate}
From these boundary conditions compatible with supersymmetry one derives the allowed discrete set and  the corresponding Fermi sea
in the $c\to +\infty$ limit. Also the behaviour in the Large $N$, Large $L$ regime may be analyzed from the ensuing spectral densities.

\section{Summary and further comments}
In this paper we have succeeded in building systems in ${\cal N}=2$ extended Supersymmetric Quantum Mechanics such that the scalar sectors
reproduce the Lieb-Liniger-Yang system of $N$ particles interacting pairwise at coinciding points through Dirac $\delta$-potentials,
either repulsive or attractive depending on the choice of superpotential. There are also matricial sectors where the wave functions are \lq\lq spinors\rq\rq of {\footnotesize $\left(\begin{array}{c} N \\ R\end{array}\right)$}=$\frac{N!}{(N-R)!R!}$ components, $R=1,2, \cdots, N-1$. The supersymmetric Hamiltonian acting in each sector becomes a {\footnotesize $\left(\begin{array}{c} N \\ R\end{array}\right)\times \left(\begin{array}{c} N \\ R\end{array}\right)$} matrix operator. There are Schr$\ddot{\rm o}$dinger operators along the diagonal with $\delta$-point interactions but
more contact $\delta$-interactions arise off the main diagonal. For any number of particles there is one bosonic and one fermionic ground state: supersymmetry is not spontaneously boken but the Witten index is zero.

One may understand the Calogero-Moser-Sutherland model with pair-wise potentials given by the square of the hyperbolic secant as built from the one-particle reflectionless P$\ddot{\rm o}$sch-Teller Schr$\ddot{\rm o}$dinger operator. It is conceivable that a similar generalization of the potentials described in Reference \cite{Arancibia} will lead to interesting one-dimensional $N$-body systems. The same prospect can be extracted from the one-particle potentials derived in Reference \cite{Plyushchay} concerning their promotion to pair-wise potentials
in order to generalize the elliptic (Weierstrass) CMS $N$-body system. Finally, the promotion to a supersymmetric status of the $(1+1)$-dimensional scalar field theories under the influence of $\delta$ and $\delta^\prime$ external backgrounds, see References \cite{Mosquera} and \cite{Guilarte}, may be a good scenario where contributions of bosonic and fermionic fluctuations to the Casimir effect can be compared.


\begin{thebibliography}{A}

\bibitem {Bethe}H. A. Bethe, \textit{Zur Theorie der Metalle, I. Eigenwerte und Eigenfunktionen der linearen
Atomketten}, Zeits. Phys. \textbf{71}, (1931), 205-226; English translation \textit{On the theory of metals, I. Eigenvalues and eigenfunctions of a linear chain of atoms}, in Bethe, H.A., Selected Works of H. A. Bethe With Commentary, World Scientific, Singapore, 1996, pgs. 689-716.

\bibitem {Hulthen}L. Hulth\'{e}n, \textit{\"Uber das austauschproblem eines kristalles}, Ark. Mat. Astron. Fys. A \textbf{26}, (1938), 1-106.

\bibitem {Orbach}R. Orbach, \textit{Linear antiferromagnetic chain with anisotropic coupling}, Phys. Rev., \textbf{112}, (1958), 309-316.

\bibitem {Walker}L. R. Walker, \textit{Antiferromagnetic linear chain}, Phys. Rev., \textbf{116}, (1959), 1089-1090.

\bibitem {YYan}C. N. Yang and C. P. Yang, \textit{One-dimensional chain of anisotropic spin-spin interactions. I, Proof of Bethe's hypothesis for the ground state in a finite system}, Phys. Rev. \textbf{150}, (1966), 321-327.

\bibitem {YYang}C. N. Yang and C. P. Yang, \textit{One-dimensional chain of anisotropic spin-spin interactions. I, Properties of the ground state energy per lattice site for an infinite system}, Phys. Rev. \textbf{150}, (1966), 327-339.

\bibitem {Lieb}E. Lieb, W. Liniger, \emph{Exact analysis of an interacting Bose gas I: The general solution and the ground state}, Phys. Rev. \textbf{130} No. 4 (1963) 1605-1616.

\bibitem {Liniger}E. H. Lieb, \textit{Exact analysis of an interacting Bose gas II: The excitation spectrum}, Phys. Rev., \textbf{130}, No. 4 (1963), 1616-1624.

\bibitem {McGuire}J. B. McGuire, \textit{Study of exactly soluble one-dimensional N-body problems}, J. Math. Phys., \textbf{5}, No. 5 (1964), 622-636.

\bibitem {Yang}C. N. Yang, \textit{Some exact results for the many-body problem in one dimension with repulsive delta-function interaction}, Phys. Rev. Lett., \textbf{19}, (1967), 1312-1314.

\bibitem {Baxt}R. J. Baxter, \textit{Generalized ferroelectric model on a square lattice}, Stud. Appl. Math., \textbf{50} (1971) 51-69.

\bibitem {Baxte}R. J. Baxter, \textit{Partition function of the eight-vertex lattice model}, Ann. Phys. (NY), \textbf{70} (1972) 193-228.

\bibitem {Baxter}R. J. Baxter, \textit{One-Dimensional Anisotropic Heisenberg Chain}, Ann. Phys. (NY), \textbf{70} (1972) 323-337.

\bibitem {Wu}F. Y. Wu, \textit{Knot theory and statistical mechanics}, Rev. Mod. Phys. \textbf{64}, (1992), 1099-1131.

\bibitem {Wadati}M. Wadati, \textit{Knot theory and integrable systems}, Springer series in non-linear dynamics, (1993), 468-486.

\bibitem {Sklyanin}E. K. Sklyanin and L. D. Faddeev, \textit{Quantum mechanical approach to completely integrable models of field theory}, Dokl. Akad. Nauk SSSR , \textbf{243}, No. 6, (1978), 1430-1433.

\bibitem {ESklyanin}E. K. Sklyanin and L. D. Faddeev, \textit{Method of the inverse scattering problem and quantum nonlinear Schr\"odinger
equation}, Dokl. Akad. Nauk SSSR , \textbf{244}, No. 6, (1978), 1337-1341.

\bibitem {Zamolodchikov}Alexander B. Zamolodchikov and Aleksey B. Zamolodchikov, \textit{Factorized S-matrices in two dimensions as the exact solutions of certain relativistic quantum field theory models}, Ann. Phys. \textbf{120}, (1979), 253-291.


\bibitem {Takahashi}M. Takahashi, \textit{Thermodynamics of one-dimensional solvable models}, Cambridge University Press, Cambridge, 1999.

\bibitem{Ushveridze}A. G. Ushveridze, \emph{Analytic properties of energy levels in models with delta-function potentials}, J. Phys. A: Math. Gen. \textbf{21} (1988) 955-970.

\bibitem {Ladislav}L. \v{S}amaj, \textit{Introduction to integrable many-body systems I}, Acta Physica Slovaca, \textbf{58}, No. 6 (2008), 811-946.

\bibitem {Kundu}A. Kundu, \textit{Quantum Integrable Systems: Construction, Solution, Algebraic Aspect}, [arXiv:hep-th/9612046v1].

\bibitem {Jimbo}M. Jimbo, ed. by, \textit{Yang-Baxter equations in integrable systems}, Advanced series in mathematical physics, v. 10, World Scientific.

\bibitem {Hallnas}M. Halln\"as, \textit{Quantum many-body systems exactly solved by special functions}, Scientific thesis for the degree of Doctor of Philosophy (PhD) in the subject area of Theoretical Physics, Royal Institute of Technology (KTH) Albanova University Center SE-106 91, Stockholm, Sweden (2007).

\bibitem {Sutherland}B. Sutherland, \textit{Beautiful Models: 70 Years of Exactly Solved Quantum Many-Body Problems}, World Scientific Publishing Co., 2004.

\bibitem{Calogero0}F. Calogero, \textit{Solution of one-dimensional $N$-body problems with quadratic and/or inversely quadratic pair of potentials}, J. Math. Phys. \textbf{12} (1971) 419-433

\bibitem {Calogero}F. Calogero and A. Degasperis, \textit{Comparison between the exact and Hartree solutions of a one-dimensional many-body problem}, Phys. Rev. A \textbf{11}, (1975), 265.

\bibitem{Freedman} D. Z. Freedman and P. F. Mende, \textit{An exactly solvable $N$ particle system in supersymmetric quantum mechanics}, Nuc. Phys. \textbf{B344} (1999) 317-343

\bibitem{Brink0} L. Brink, T. H. Hansson, S. Konstein, and M. A., Vasiliev, \textit{The Calogero model: Anyonic representation, fermionic extension and supersymmetry} Nucl. Phys. \textbf{B401} (1993) 591-612

\bibitem{Shastry} B. S. Shastry and B. Sutherland, \textit{Superlax pairs and infinite symmetries in the $\frac{1}{r^2}$ system} Phys. Rev. Lett. \textbf{70} (1993) 4029-4033

\bibitem{Brink}L. Brink, A. Turbiner, N. Wyllard, \textit{Hidden algebras of the (super) Calogero and Sutherland models}, J. Math. Phys. \textbf{39}(1998) 1285-1315.

\bibitem{Desrosier}P. Desrosiers, L. Lapointe, and P. Mathieu, \textit{Supersymmetric  Calogero-Moser-Sutherland models and Jack superpolynomials}, Proceedings of the Workshop on superintegrability in classical and quantum systems, CRM 2002, ed. P. Winternitz.

\bibitem{Manton} A. J. Bordner, N. S. Manton, and R. Sasaki, \textit{Calogero-Moser models V: supersymmetry and quantum Lax pair}, Progr. Theor. Phys. \textbf{103} (2000) 463-487

\bibitem{Wipf} A. Kirchberg, J. D. L\"ange, P. A. G. Pisani, and A. Wipf, \textit{Algebraic solution of the supersymmetric Hydrogen atom in $d$ dimensions}, Ann. Phys. \textbf{303} (2003) 359-388

\bibitem {Freund} P. G. O. Freund, \textit{Introduction to supersymmetry}, Cambridge University Press, Cambridge, 1986.

\bibitem {Matgui} A. A. Izquierdo, M. A. Gonzalez Leon, M. de la Torre Mayado and J. Mateos Guilarte, \textit{On two-dimensional superpotentials: from classical Hamilton-Jacobi theory to 2D supersymmetric quantum mechanics}, J. Phys. A: Math. Gen. \textbf{37} (2004), 10323-10338.
\bibitem{Mosquera1} J. M. Mu$\tilde{\rm n}$oz Casta$\tilde{\rm n}$eda, J. Mateos Guilarte, and A. Moreno Mosquera, \textit{On supersymmetric Dirac Delta interactions}, Eur. Phys. J. Plus \textbf{130} (2015) 3, 48

\bibitem{Arancibia} A. Arancibia, J. Mateos Guilarte,  M. S. Plyushchay, \textit{The effect of scalings and translations in the supersymmetric structure of soliton systems}, Phys. Rev. \textbf{D 87}, 045009 (2013)

\bibitem{Plyushchay} A. Arancibia, F. Correa, V. Jakubsky, J. Mateos Guilarte, and M. S. Plyushchay, \textit{Soliton defects in one-gap periodic system and exotic supersymmetry}, Phys. Rev. \textbf{D 90} (2014) 125041

\bibitem{Mosquera} J. M. Mu$\tilde{\rm n}$oz Casta$\tilde{\rm n}$eda, J. Mateos Guilarte, and A. Moreno Mosquera, \textit{Quantum vacuum energies and Casimir forces between partially transparent $\delta$-function plates}, Phys. Rev. \textbf{D 87} (2013) 105020

\bibitem{Guilarte} J. M. Mu$\tilde{\rm n}$oz Casta$\tilde{\rm n}$eda, and J. Mateos Guilarte, \textit{$\delta$-$\delta^\prime$ generalized Robin boundary conditions and quantum vacuum fluctuations}, Phys. Rev. \textbf{D 91} (2015) 025028

\end{thebibliography}
\end{document}